\documentclass[aps,prb,twocolumn,superscriptaddress,amsmath,amssymb,floatfix]{revtex4-2}

\usepackage{amsmath,amssymb,amsfonts}      
\usepackage{graphicx}                      
\usepackage{dcolumn}                      
\usepackage{bm}                           
\usepackage{hyperref}                     
\usepackage[usenames,dvipsnames]{xcolor}  
\usepackage{booktabs}                     
\usepackage{siunitx}                      
\usepackage{mhchem}                  
\usepackage[T1]{fontenc}                  
\usepackage{rotating}
\usepackage[toc,page]{appendix}
\usepackage{multirow}

\usepackage{placeins} 

\hypersetup{
  colorlinks=true,
  linkcolor=BlueViolet,
  citecolor=ForestGreen,
  urlcolor=MidnightBlue
}
\begin{document}

\title{Absence of Long‑Range Order and Magnetic Anisotropy in the Triangular Magnet  NdMgAl\textsubscript{11}O\textsubscript{19}}

\author{Sonu Kumar}
\email{sonu.kumar@matfyz.cuni.cz}
\affiliation{Charles University, Faculty of Mathematics and Physics, Department of Condensed Matter Physics, Prague, Czech Republic}
\affiliation{Adam Mickiewicz University, Faculty of Physics and Astronomy, Department of Experimental Physics of Condensed Phase, Poznan, Poland}

\author{Gaël Bastien}
\affiliation{Charles University, Faculty of Mathematics and Physics, Department of Condensed Matter Physics, Prague, Czech Republic}

\author{Jan Prokleška}
\affiliation{Charles University, Faculty of Mathematics and Physics, Department of Condensed Matter Physics, Prague, Czech Republic}

\author{Mateusz Kempiński}
\affiliation{Adam Mickiewicz University, Faculty of Physics and Astronomy, Department of Experimental Physics of Condensed Phase, Poznan, Poland}

\author{Wojciech Kempiński}
\affiliation{Institute of Molecular Physics, Polish Academy of Sciences
Department of Low Temperature Physics, Quantum Materials and Technologies, Poznan, Poland.}

\author{Karol Załęski}
\affiliation{Adam Mickiewicz University, NanoBioMedical Centre, Poznan, Poland}

\author{Andrej Kancko}
\affiliation{Charles University, Faculty of Mathematics and Physics, Department of Condensed Matter Physics, Prague, Czech Republic}

\author{Cinthia Antunes Corr\^ea}
\affiliation{Institute of Physics of the Czech Academy of Sciences, Na Slovance, Prague, Czech Republic}

\author{T. Treu}
\affiliation{Experimental Physics VI, Center for Electronic Correlations and Magnetism, Institute of Physics,
University of Augsburg, 86159 Augsburg, Germany}

\author{P. Gegenwart}
\affiliation{Experimental Physics VI, Center for Electronic Correlations and Magnetism, Institute of Physics,
University of Augsburg, 86159 Augsburg, Germany}

\author{Ma\l{}gorzata Śliwińska-Bartkowiak}
\affiliation{Adam Mickiewicz University, Faculty of Physics and Astronomy, Department of Experimental Physics of Condensed Phase, Poznan, Poland}

\author{Ross H. Colman}
\email{ross.colman@matfyz.cuni.cz}
\affiliation{Charles University, Faculty of Mathematics and Physics, Department of Condensed Matter Physics, Prague, Czech Republic}

\begin{abstract}
We investigated the rare-earth triangular-lattice antiferromagnet NdMgAl$_{11}$O$_{19}$ using single-crystal magnetization (1.8~K~$\leq~T~\leq$~300~K, $\mu_0H \leq 7$~T) and specific-heat measurements down to 45~mK. The dc susceptibility confirms a well-isolated Kramers doublet ground state with pronounced Ising-type anisotropy, with $g_{c} \approx 3.7$, $g_{ab} \approx 1.45$. Curie--Weiss fits yield weak, anisotropic antiferromagnetic exchange, with $\theta_{c} = -0.54$~K and $\theta_{ab} = -0.87$~K. Heat capacity measurements show no long-range magnetic order down to 40~mK, corresponding to a frustration index $f \gtrsim 20$. Instead, $C_m/T$ exhibits a broad maximum near 0.081~K whose magnitude and field evolution are consistent with short-range correlations in an anisotropic triangular lattice. Applied magnetic fields open a Zeeman gap where the specific-heat anomaly follows $\Delta = g \mu_B \mu_0 H$, and $M(H,T)$ is well described by a Brillouin function for an effective $J = \tfrac{1}{2}$ moment. The field tuning of the low-temperature entropy manifold allows self-cooling from 1.8 K to 53 mK by adiabatic demagnetisation from a 9T field. These results identify NdMgAl$_{11}$O$_{19}$ as a nearly ideal weak-exchange triangular magnet with a field-tunable correlated ground state, where two-dimensional crossover effects may emerge from frustrated XXZ interactions.
\end{abstract}

\keywords{Quantum spin liquid, Triangular lattice, NdMgAl\textsubscript{11}O\textsubscript{19}, Geometric frustration, Crystal electric field, rotational magnetocaloric}

\maketitle

\section{Introduction}

Geometrically frustrated magnets on triangular lattices often exhibit unconventional magnetic states, including quantum spin liquids (QSLs), owing to the inherent competition among spins that prevents classical long-range order \cite{Ramirez2025,Balents2010,Savary2017}. In such systems, pronounced quantum fluctuations can lead to anomalous thermodynamic responses, persistent spin dynamics, or partially ordered phases. The suppression of magnetic order through frustration has proven technologically important with several frustrated magnets proving useful for adiabatic demagnetisation refrigerant (ADR)~\cite{Tokiwa2021,Treu2025}. Rare-earth triangular-lattice magnets are especially interesting because strong spin-orbit coupling (SOC) and crystal electric field (CEF) effects frequently yield large anisotropy and unusual correlation phenomena \cite{Qin2021,Xie2024}.

Among rare-earth triangular-lattice systems, Yb- and Tm-based materials have received extensive focus \cite{Li2015,Voma2021,Somesh2023,Khatua2024,Shen2019,Treu2025}. YbMgGaO$_4$ stirred significant debate over potential gapless QSL behavior and the role of site disorder \cite{Rao2021,Kimchi2018,Li2015,Li2017}. TmMgGaO$_4$ revealed essential insights into triangular-lattice quantum Ising physics and complex topological transitions \cite{Shen2019,Li2020,Huang2024,Liu2020,Li2020}. Kramers ions (Nd$^{3+}$, Yb$^{3+}$, Ce$^{3+}$), which possess half-integer total angular momentum $J$, are guaranteed to host at least a doubly degenerate ground state due to time-reversal symmetry. In contrast, non-Kramers ions (Tm$^{3+}$, Pr$^{3+}$) with integer $J$ may have nondegenerate or quasi-degenerate ground states, depending sensitively on the crystal electric field environment. This distinction leads to qualitatively different low-temperature behaviors, especially in the presence of disorder or quantum fluctuations.
 This anisotropy often manifests in Ising-type or XY-type characters, thereby stabilizing novel partially disordered states, spin-ice-like phases, field induced magnetic super-solid phases, or non-trivial excitations with strong quantum fluctuations \cite{Bramwell2020,Seabra2011}.

In recent years, the rare-earth hexaaluminate family Ln(Mg/Zn)Al$_{11}$O$_{19}$ (Ln = lanthanide) has emerged as a model platform for frustrated triangular-lattice magnetism \cite{Ma2024,Kumar2025,Bu2022,gao2024,Bastien2025,Kumar2025PrMgAl11O19ME,Kumar2026SmMgAl11O19,Kumar2025CeMgAl11O19ME,cao2025}. 
Across the series, the interplay of geometric frustration, SOC, and crystal-field anisotropy stabilizes a range of unconventional low-temperature states: 
PrMgAl$_{11}$O$_{19}$ hosts a disorder-stabilized quantum Ising–like state with induced transverse field \cite{Kumar2025,Ma2024}, 
CeMgAl$_{11}$O$_{19}$ lies proximate to an XXZ-type quantum critical point and exhibits a broad excitation continuum \cite{gao2024,Bastien2025}, 
while the magnetic response of SmMgAl$_{11}$O$_{19}$ is strongly quenched \cite{Kumar2026SmMgAl11O19}.
 However, the properties of NdMgAl$_{11}$O$_{19}$ remain less explored, despite Nd$^{3+}$ being a Kramers ion that often displays notable low-temperature anisotropy and collective phenomena \cite{Ashtar2019,Soderholm1991,Chamorro2023,Arh2022}.

This work focuses on the novel rare-earth triangular lattice compound NdMgAl$_{11}$O$_{19}$. The Nd$^{3+}$ ion has a total angular momentum of $J=9/2$, which, in the presence of crystal-field splitting, often results in a low-lying Kramers doublet, significantly influencing magnetic anisotropy at low temperatures \cite{Arh2022, cao2025}. Our motivation is to explore how weak antiferromagnetic interactions, $g$-factor anisotropy, and geometric frustration collectively shape the unusual magnetic and thermodynamic responses observed in NdMgAl$_{11}$O$_{19}$, as well as highlighting its magnetocaloric refrigeration potential. We compare and contrast our single crystal study to the polycrystal study of the isostructural analogue NdZnAl$_{11}$O$_{19}$~\cite{cao2025}.

\section{Methods}

The synthesis and subsequent crystal growth of NdMgAl$_{11}$O$_{19}$ were successfully carried out using a solid-state reaction and the optical floating zone (OFZ) method. Initial precursor binary oxides (Nd$_6$O$_{11}$, MgO, and Al$_2$O$_3$ of \SI{99.99}{\%} purity) were calcined at \SI{800}{\celsius} for \SI{24}{hours} to remove moisture or carbonate contamination. Following calcination, the oxides were weighed in stoichiometric ratio, thoroughly mixed, and ground to ensure homogeneity. The mixture was pressed into cylindrical rods with dimensions of \SI{6}{mm} in diameter and \SI{100}{mm} in length. Densification was achieved using a quasi-hydrostatic pressure of \SI{2}{tons} for \SI{15}{minutes}. The rods were sintered in air at \SI{1200}{\celsius} for \SI{72}{hours} to promote solid-state reaction and improve density. Growth was performed under an air atmosphere with a slight overpressure of \SI{1}{atm} and a flow rate of \SI{3}{L/min}. The sintered rods were used as both feed and seed material for crystal growth. The feed and seed rods were counter-rotated at \SI{30}{rpm} to improve temperature distribution and material mixing in the molten zone. The growth rate was maintained at \SI{2}{mm/h}. Further details about the synthesis method are reported elsewhere \cite{Kumar2025,Bastien2025}.

The growth resulted in a light purple ingot containing multiple grains, with visible grain boundaries. Individual grains were separated mechanically using a wire saw and gentle cleavage, taking advantage of the strong natural cleavage parallel to the \textit{ab}-plane. The single-crystalline nature of the resulting pieces was confirmed using backscattered Laue X-ray diffraction.
 Single-crystal X-ray diffraction (SCXRD) measurements were carried out at \SI{95}{K} using a Rigaku SuperNova diffractometer equipped with an Atlas S2 CCD detector and a mirror-collimated Mo K$\alpha$ radiation source ($\lambda = \SI{0.71073}{\angstrom}$) from a micro-focus sealed tube. Data integration, scaling, and absorption correction were performed with CrysAlis Pro \cite{crysalispro}, incorporating an empirical absorption correction based on spherical harmonics \cite{clark1995} and an analytical numerical correction using Gaussian integration over a multifaceted crystal model via the SCALE3 ABSPACK scaling algorithm. The crystal structure was solved by the charge-flipping method using Superflip \cite{palatinus2007} and refined through full-matrix least-squares on $F^2$ using Jana2020 \cite{petricek2023}.

The DC magnetic susceptibility was measured using a Quantum Design SQUID Magnetic Property Measurement System (MPMS). Specific heat was measured from room temperature down to \SI{40}{mK}, using a combination of the Quantum Design Physical Property Measurement System (PPMS), the PPMS $^3$He option, and an Oxford Instruments Triton dilution refrigerator. Multiple pieces of single crystals from two growths were used for the bulk measurements, with consistent results confirming the sample quality and reproducibility of response. This report shows results obtained from pieces of \SI{8.87}{mg} and \SI{4.39}{mg}. LaMgAl$_{11}$O$_{19}$ single crystals were also grown in air using the floating zone method, and their specific heat measurement was used to isolate the magnetic contribution ($C_m$) of the Nd counterpart. Crystals of LaMgAl$_{11}$O$_{19}$ contain a tiny amount of magnetic impurities influencing the specific heat at low temperatures; therefore, in the low-temperature limit ($T<$ \SI{1}{K}), the phononic background was obtained from a $T^3$ extrapolation of the measured specific heat of LaMgAl$_{11}$O$_{19}$.

The measurement of cooling potential by adiabatic demagnetization refrigeration (ADR) involved the self-cooling of the sample from 1.8\,K by demagnetization from 9 T to zero applied field over a period of 30 minutes quasi-adiabatically, on a pressed disk of powdered sample (1~g) intimately mixed with silver powder (1~g) for enhanced thermal conductivity, within a Quantum Design PPMS (For more information, see~\cite{Tokiwa2021, Treu2025,Klinger2025}).

Electron paramagnetic resonance (EPR) spectra were acquired using a Radiopan ES/X spectrometer equipped with an Oxford Instruments helium-flow cryostat, operating over \SIrange{1.5}{300}{\kelvin}.
Single-crystal fragments were mounted on a rotating quartz holder to probe signal anisotropy.

\section{Results}

\subsection{Structural Characterization and point-charge CEF Calculations}

\begin{figure*}[t]
  \includegraphics[width=\textwidth]{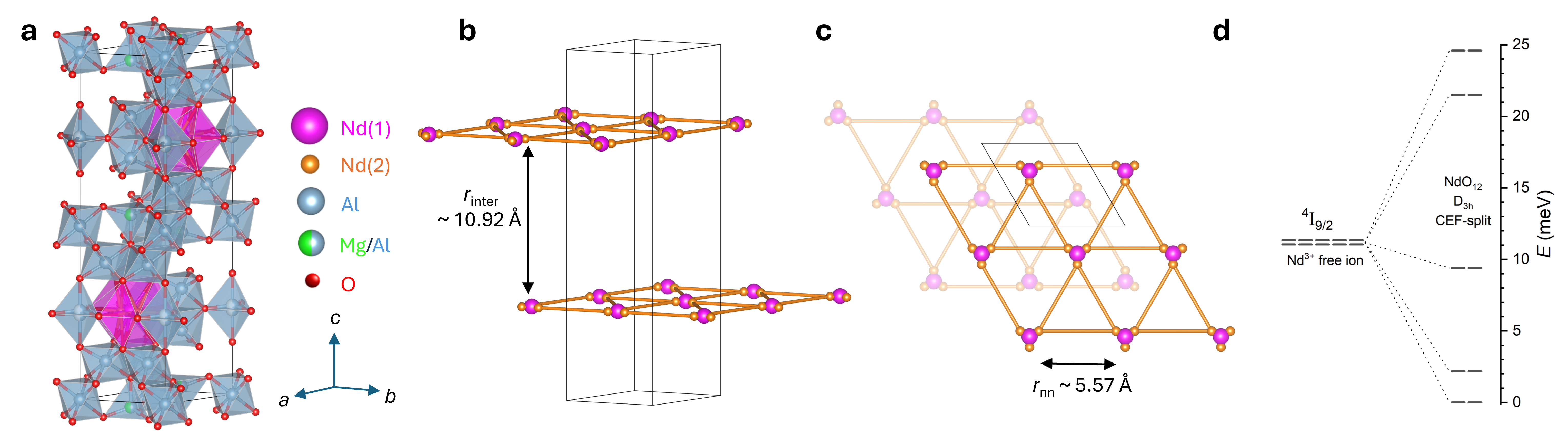}
  \caption{The magnetoplumbite structure of NdMgAl$_{11}$O$_{19}$ (a), with magnetic Nd$^{3+}$ triangular lattice inter-planar (b) and intra-planar (c) separation distances. (d) A representation of the point-charge-approximation calculation of crystal electric field splitting of the Nd$^{3+}$  $^4$I$_{9/2}$ free ion electronic levels, resulting in a Kramers' doublet ground-state, separated from the first excited doublet level by 2.174 meV (25.2 K).}
  \label{fig:Structure}
\end{figure*}

Our single-crystal X-ray diffraction measurements confirm that NdMgAl$_{11}$O$_{19}$ adopts the expected magnetoplumbite structure, consistent with previous reports on related LaMgAl$_{11}$O$_{19}$ compounds (La = lanthanides), seen in Figure~\ref{fig:Structure}~\cite{Bastien2025,Kumar2025,Li2024,gao2024,Saber1981,Gasperin1984}. In this structure, Nd$^{3+}$ ions occupy 12-fold coordinated oxygen
cages arranged in triangular layers, with each NdO$_{12}$ polyhedron sharing a single in-plane oxygen vertex with its neighbors. These magnetic layers are separated by approximately \SI{11}{\angstrom} along the $c$-axis by a nonmagnetic slab of edge- and face-sharing Al and Mg polyhedra, forming a spinel-like block.

The single-crystal structure, refined at \SI{95}{K}, is in excellent agreement with those of the Pr and Ce analogues and is provided in the Supplementary Information \cite{SI}. As with other members of this family, four sources of inherent local structural disorder are found possible: (i) a mixed Mg/Al occupancy at the Al(4)/Mg(1) site, located far from the Nd layers and having negligible magnetic impact; (ii) a positionally disordered Al(5) ion residing in a double-well potential within the bipyramidal oxygen cage, off-centered along $c$ by $\delta \approx \SI{0.31}{\angstrom}$; (iii) a positional disorder of the Nd$^{3+}$ ion within its 12-fold cage, described below; and (iv) a refined incomplete occupancy of the Nd site(s).

Refinement quality was significantly improved by incorporating a second Nd site displaced by \SI{0.83}{\angstrom} from the high-symmetry 2d Wyckoff position, leading to refined occupancies of 0.852(4):0.0188(9) for Nd1$_{2d}$:Nd2$_{6h}$, respectively. This off-center site, situated on a 6h Wyckoff position, is supported by residual electron density maxima in Fourier difference maps and aligns with similar positional disorder reported in PrMgAl$_{11}$O$_{19}$ \cite{Kumar2025,Cao2024}. The final composition Nd$_{0.908}$MgAl$_{11}$O$_{19}$ suggests a \SI{9.2(4)}{\%} total Nd$^{3+}$ deficiency, also commonly seen in several other members of this family \cite{Kumar2025,Bastien2025, Cao2024, gao2024}. While such positional disorder can significantly affect the magnetic properties of non-Kramers ions due to sensitivity to shift of gaps between low-lying singlet CEF levels, its impact in the present case is expected to be much smaller. Nd$^{3+}$ is a Kramers ion and thus retains a degenerate doublet ground state regardless of local point symmetry. Moreover, our zero-field specific heat data show no low-lying Schottky anomalies (specific heat section), further indicating a well-isolated ground state Kramers doublet at temperatures below 10 K.

To further investigate the crystal field scheme, we performed point-charge CEF calculations using the \texttt{PyCrystalField} software package~\cite{Scheie2021}, considering only the majority-occupied 2d Nd$^{3+}$ Wyckoff site. Our point charge calculations take into account the first coordination sphere of the Nd$^{3+}$ ion, the NdO$_{12}$ polyhedron, so should be taken as an initial estimation only. However, they lead to the splitting of the 10-fold Nd$^{3+}$ $^4$I$_{9/2}$ levels into 5 pairs of Kramers degenerate doublet states. Full details of the calculated splittings are given in the Supplimentary Materials ~\cite{SI}, and graphically shown in Figure~\ref{fig:Structure}d, but importantly for interpretation of the low temperature magnetic behaviours these preliminary calculations show that the Kramers doublet ground-state is separated from the first excited doublet state by an energy of \SI{2.17}{meV} (\SI{25.2}{K}).  Such splittings are often seen for Nd$^{3+}$ oxides, leading to the common description of the ground-state properties resulting from an effective $S_{eff} = 1/2$ state. The anisotropic coordination within the NdO$_{12}$ polyhedron results in significant single ion anisotropy ($g_{x,y}=1.35$ vs. $g_{z}=4.45$,$g_{z}/g_{x}=3.31$, close to EPR results detailed in the Supplimentary Information \cite{SI}) perpendicular to the triangular lattice planes. Whilst these calculations provide an initial estimate only of the CEF scheme in NdMgAl$_{11}$O$_{19}$, we can compare with spectroscopic measurements of the isostructural analogue NdZnAl$_{11}$O$_{19}$~\cite{cao2025}. Inelastic neutron scattering observes dispersionless excitations $\sim$ 3.5, 9, 23.8, and 70 meV. The first three excitations appear close in energy to several predicted by our preliminary point-charge model, at $E_{\rm_{CEF}}=2.17,\;9.39,\;21.50,\;24.59~\mathrm{meV}$~\cite{SI}, however the model chosen for NdZnAl$_{11}$O$_{19}$ assigns the observed excitations to a mixture of the Nd1$_{2d}$ and Nd2$_{6h}$ sites, using only two of the excitations for determination of their  Nd1$_{2d}$ CEF scheme. Nevertheless, the conclusion of a separated, anisotropic ($g_{c}/g_{ab}\sim3.3$), ground-state doublet dominating the low-temperature physics remains consistent between analyses of these isostructural analogues.

\begin{figure*}
  \includegraphics[width=\columnwidth]{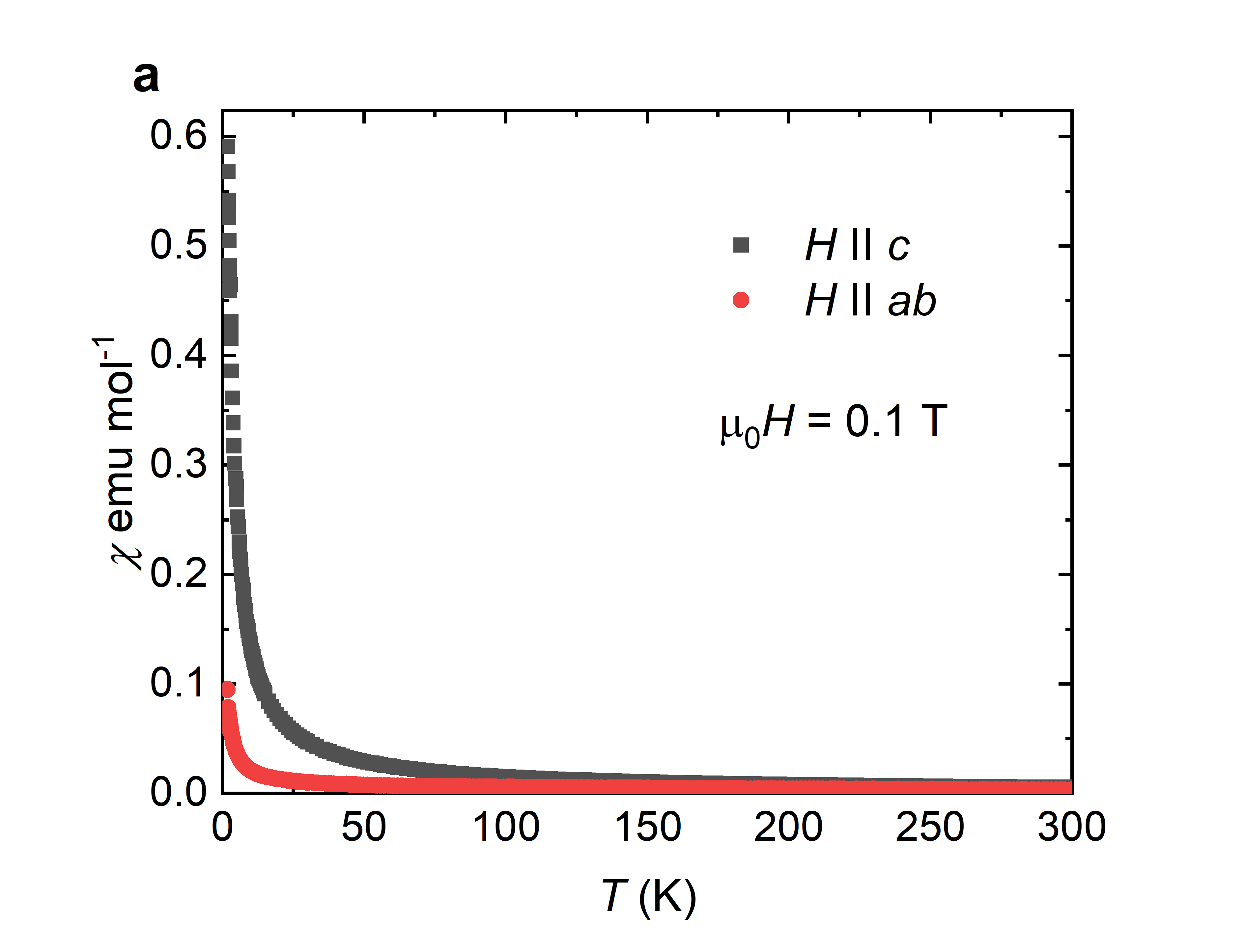}
  \includegraphics[width=\columnwidth]{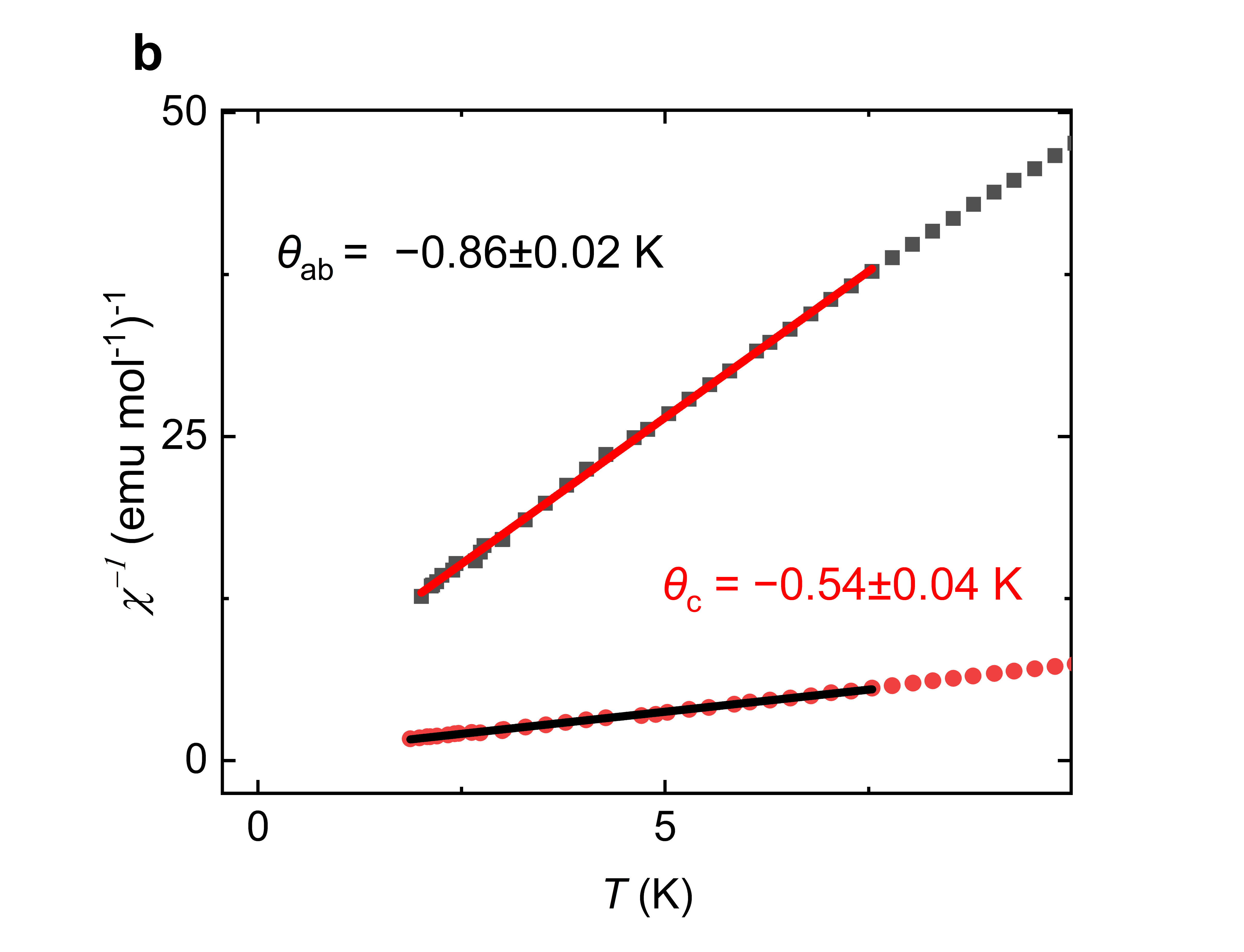}
  \caption{Magnetic susceptibility measured with an external magnetic field $\mu_oH = \SI{0.1}{T}$, applied parallel to the $c$-axis, and perpendicular to the $c$-axis (a). Corresponding inverse susceptibility with CW fits at low temperatures. The solid red lines represent the Curie-Weiss fitting of the inverse susceptibility, described in the text.}
  \label{fig:susceptibility}
\end{figure*}

Figure~\ref{fig:susceptibility} shows the magnetic susceptibility measured in a field of $\mu_0H$ = \SI{0.1}{\tesla} applied parallel to the $c$-axis, $\chi_{c}(T)$, and in the $ab$-plane, $\chi_{ab}(T)$. Down to \SI{1.8}{\kelvin} no signature of long-range magnetic order appears. At sufficiently high temperatures the Curie–Weiss law is obeyed; however, deviations from linear behavior emerge below \SI{100}{\kelvin}, likely due to changing occupation of the low-lying crystal-electric-field levels. Because the Nd$^{3+}$ ion ($J = 9/2$) is split into five Kramers doublets by the CEF, the temperature dependence of thermodynamic and magnetic properties of NdMgAl$_{11}$O$_{19}$ are strongly influenced by this level scheme (Appendix~\ref{app:cefpc}).

Fitting the linear region of $\chi^{-1}(T)$ between \SI{1.8}{\kelvin} and \SI{7.5}{\kelvin}, where we can reasonably expect significant thermal population of only the lowest lying electronic doublet (<7.5 K), yields small negative Curie–Weiss temperatures of $\theta_{\text{CW}}^{c} = -0.54(5) \si{\kelvin}$ for $H\parallel c$ and $\theta_{\text{CW}}^{ab} = -0.86(2) \si{\kelvin}$ for $H\parallel ab$, indicating weak antiferromagnetic interactions. The ratio $\chi_{c}/\chi_{ab} \approx 10$ at \SI{2}{\kelvin} reflects a strong single-ion uniaxial anisotropy, consistent with the CEF calculations, and with a predominantly $c$-axis moment, similar to that seen in CeMgAl$_{11}$O$_{19}$~\cite{Bastien2025}. The effective moment extracted from the Curie–Weiss fit is $\mu_{\mathrm{eff}} \approx 3.42\ \mu_{\mathrm{B}}$, and $\mu_{\mathrm{eff}} \approx 1.33\ \mu_{\mathrm{B}}$ corresponding to an effective $g$-factor of $g_{\mathrm{eff}} \approx 3.95$, and $g_{\mathrm{eff}} \approx 1.54$, respectively for $H\parallel c$ and $H\parallel ab$ close to that estimated by our point-charge calculations and EPR measurements (Appendix~\ref{app:cefpc} and Appendix~\ref{app:epr}).

\begin{figure*}
  \includegraphics[width=\columnwidth]{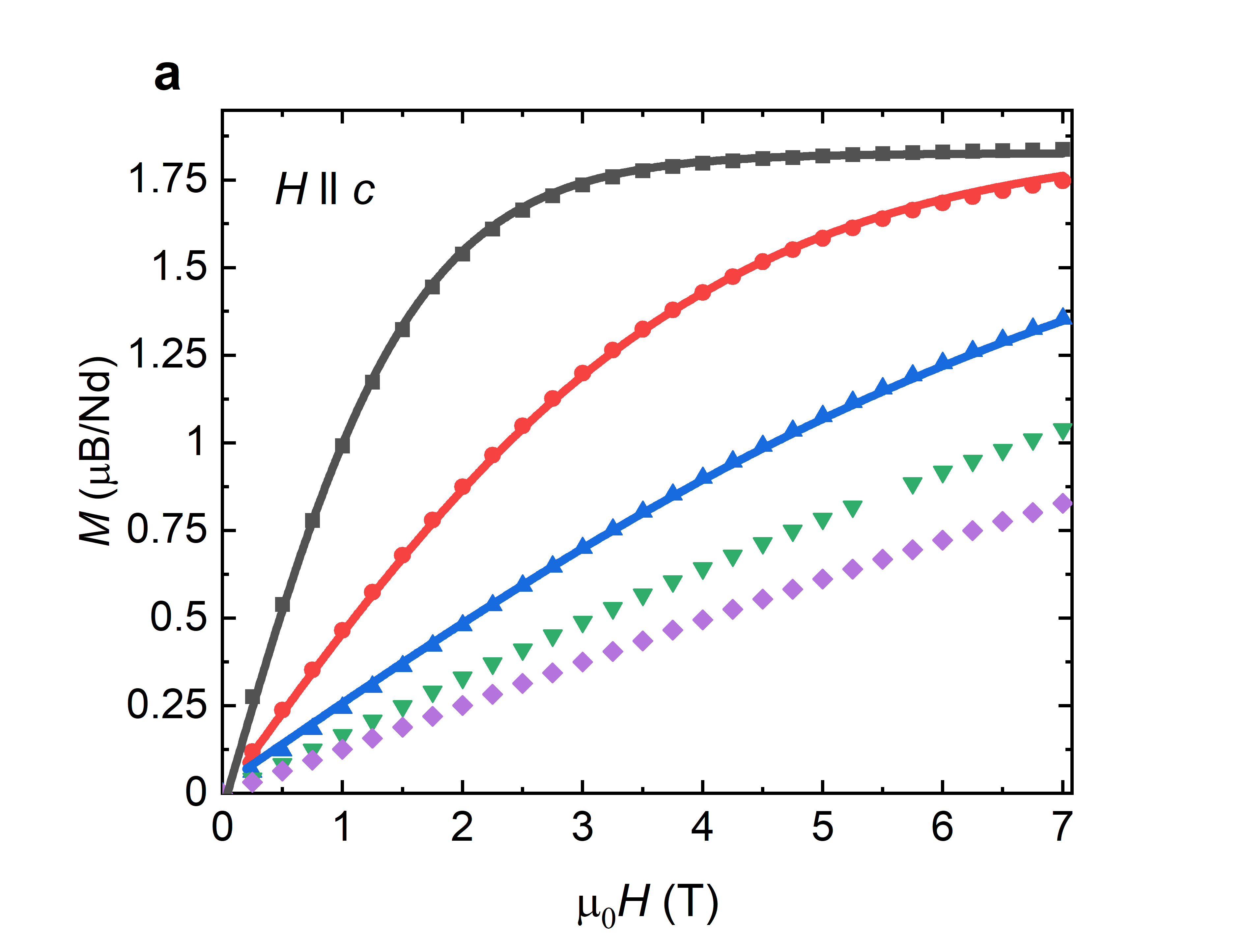}
  \includegraphics[width=\columnwidth]{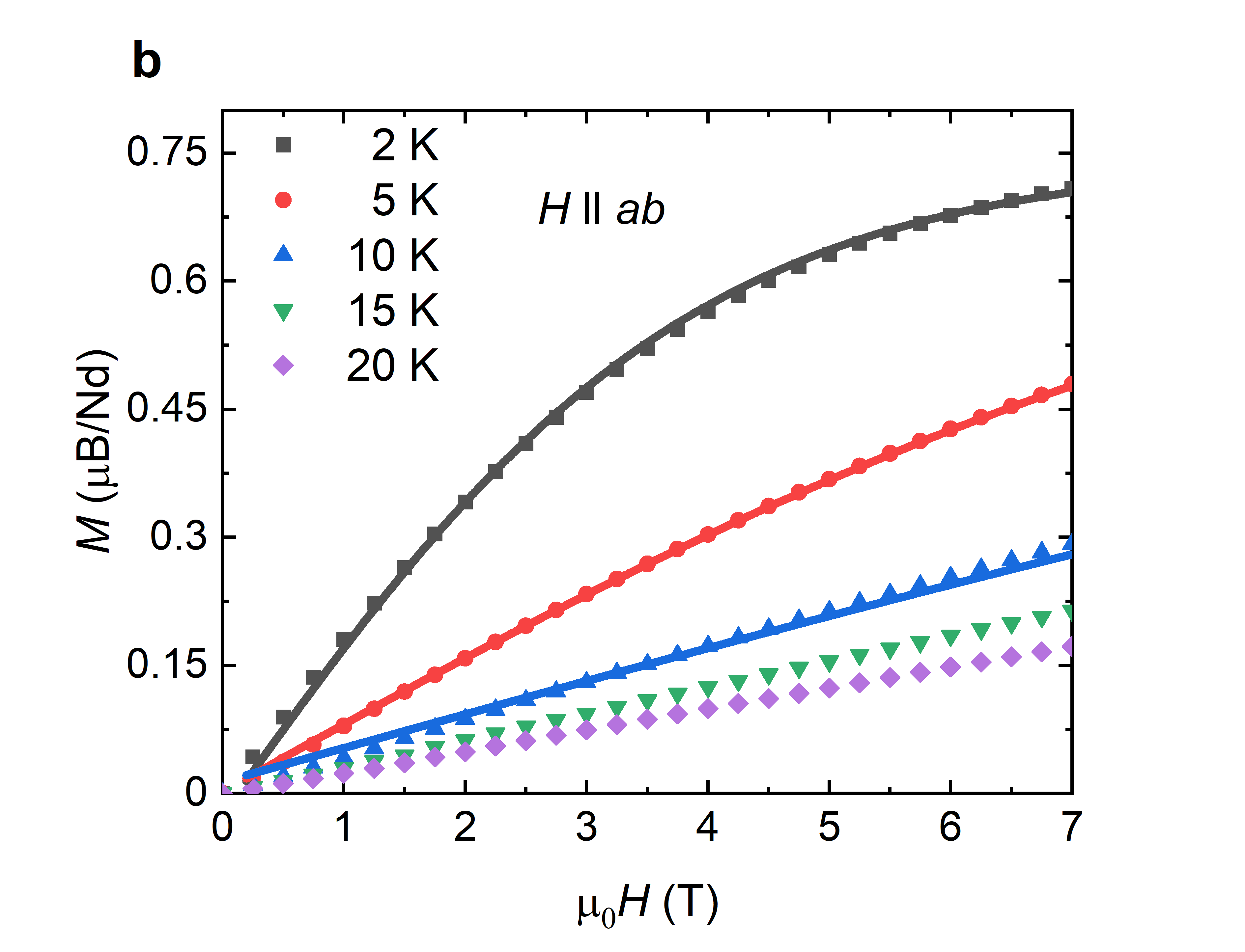} 
  \caption{Isothermal magnetization as a function of the field for fields applied along the $c$-axis (a) and in the $ab$-plane (b), with the Brillouin fit shown as solid lines.}
  \label{fig:magnetization}
\end{figure*}

Figure~\ref{fig:magnetization} shows the isothermal magnetization $M(H)$ for both field orientations. The data up to 10~K are well described by a Brillouin function for an effective $J_{\mathrm{eff}}=\tfrac{1}{2}$ doublet, supplemented by a small temperature-independent background term,
\begin{equation}
    M(H,T)=M_s\,B_J(x)+aH,
\end{equation}
where $M_s=gJ\mu_B$ and
\begin{equation}
    B_J(x)=\frac{2J+1}{2J}\coth\!\left(\frac{2J+1}{2J}x\right)
    -\frac{1}{2J}\coth\!\left(\frac{x}{2J}\right),
\end{equation}

with 
\begin{equation}
    x=\frac{gJ\mu_B H}{k_B T}.
\end{equation}

A global fit to the $T=2$, 5, and 10~K datasets (within the regime where only the ground-state doublet is significantly thermally populated) yields $g=3.75(1)$ for $H\parallel c$, with negligible $a$, confirming that the magnetization is dominated by the Kramers doublet ground state. For $H\parallel ab$, the magnetization reaches $\sim0.75~\mu_{\mathrm{B}}$/Nd at $\mu_0H=7$~T without full saturation; the corresponding global fit gives $g=1.54(1)$ and an expected saturation moment of $\approx0.77~\mu_{\mathrm{B}}$/Nd. The excellent agreement between the experimental data and the Brillouin model in both field orientations demonstrates that, for $10~K \gtrsim T\gtrsim 2$~K, the magnetization is governed by weakly interacting effective spin-$\tfrac{1}{2}$ moments, with exchange interactions becoming relevant only at lower temperatures.

\subsection{Specific Heat}

\begin{figure*}[t]
  \centering
  \includegraphics[width=0.49\textwidth]{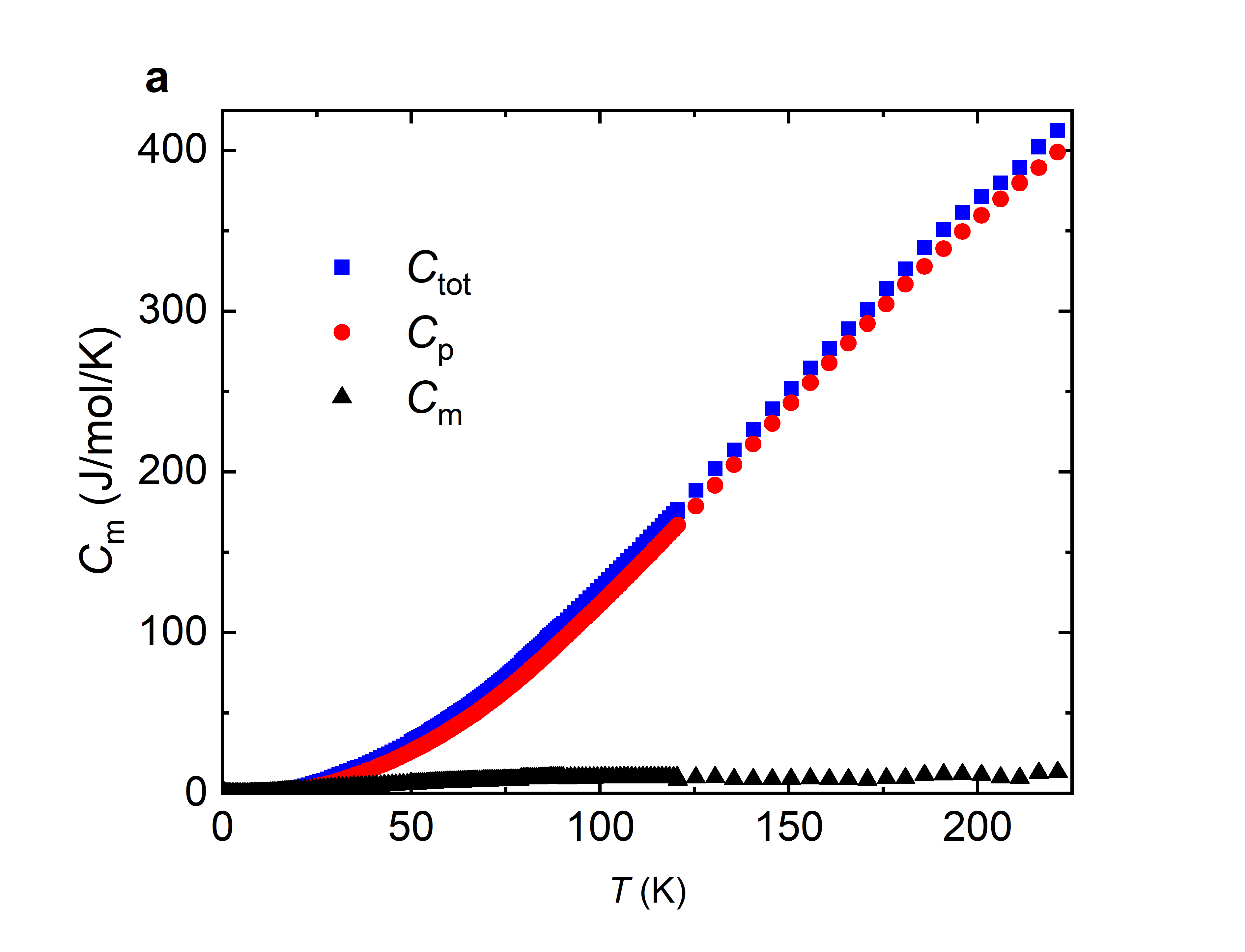}\hfill
  \includegraphics[width=0.49\textwidth]{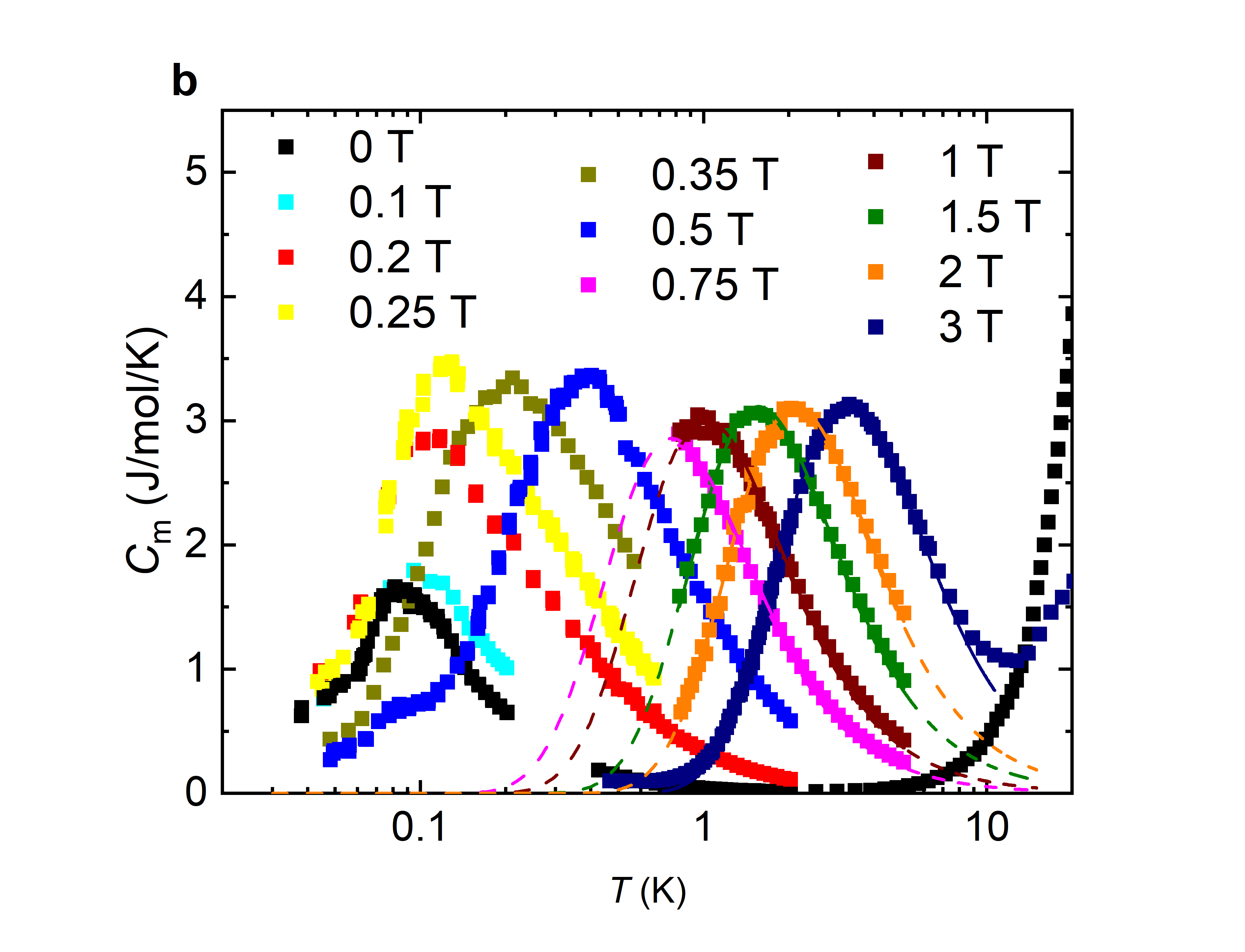}

  \vspace{2mm}

  \includegraphics[width=0.49\textwidth]{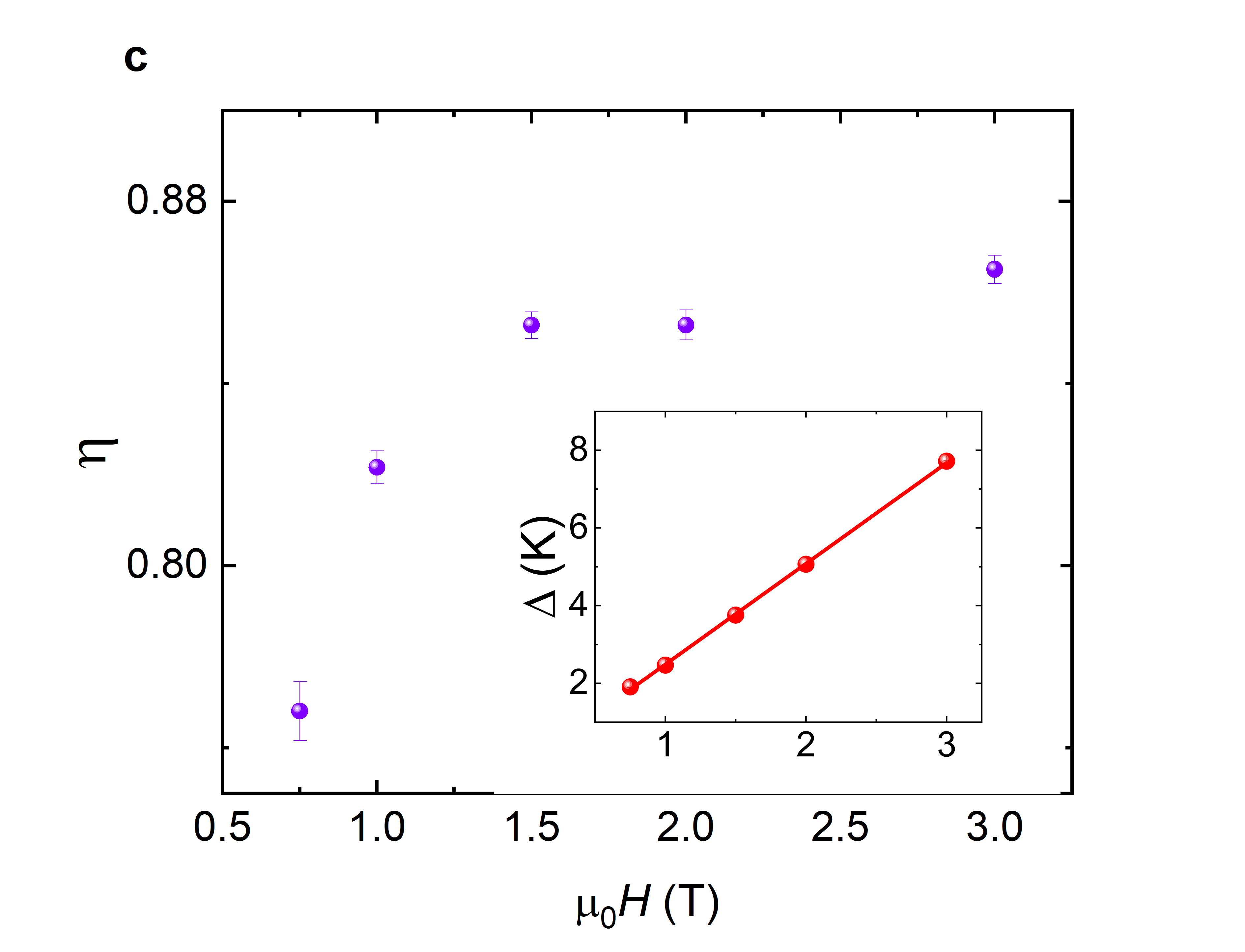}\hfill
  \includegraphics[width=0.49\textwidth]{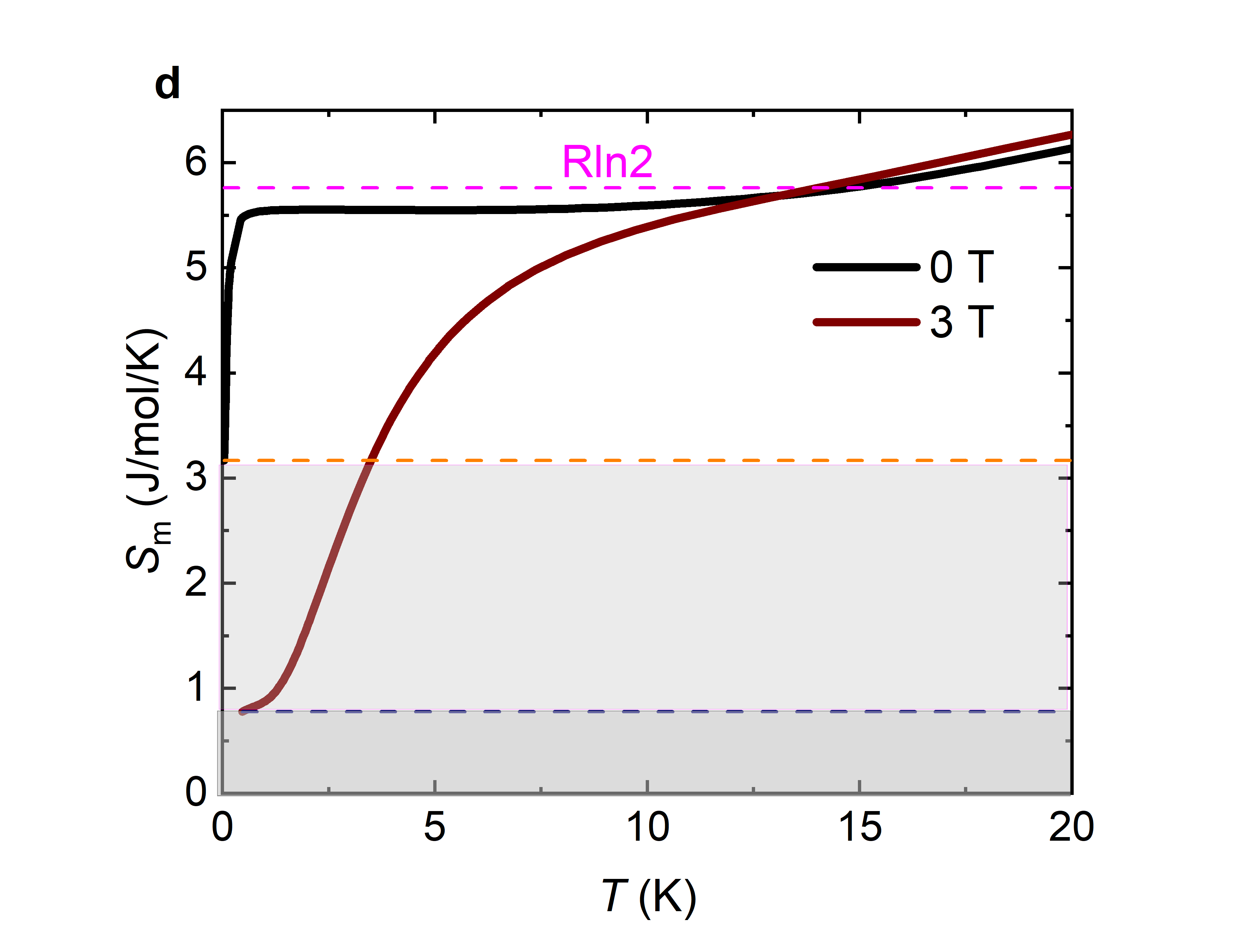}
\caption{(a) Total specific heat ($C_{\mathrm{tot}}$), non-magnetic (phonon) contribution ($C_p$), and the resulting magnetic specific heat ($C_m$) obtained by subtraction, shown as functions of temperature in zero magnetic field. (b) Magnetic specific heat $C_m(T)$ measured in various magnetic fields. Solid lines show fits to the two-level Schottky model, while dashed lines indicate extrapolated curves inferred from the Schottky analysis. (c) Field dependence of $\eta$, the fraction of ions contributing to the Schottky anomaly; inset: extracted Schottky gap $\Delta$ as a function of magnetic field. (d) Magnetic entropy $S_m(T)$. The semi-transparent gray shaded area below the dashed magenta line highlights the entropy deficit attributed to Nd deficiency, while the transparent gray shaded area between the dashed orange and dashed magenta lines marks the residual entropy remaining below $40$~mK.
}
  \label{fig:specific_heat}
\end{figure*}

The specific heat $C(T)$ of NdMgAl$_{11}$O$_{19}$ was measured down to \SI{40}{mK}. No sharp $\lambda$-type
anomaly is observed, excluding a conventional long-range ordering transition down to the base temperature
[Fig.~\ref{fig:specific_heat}(a)]. In zero field, $C(T)$ exhibits a broad hump with a maximum at
$T^{\ast}=81(2)$~mK. To isolate the magnetic contribution, we subtract the phonon background using the
non-magnetic analog LaMgAl$_{11}$O$_{19}$, defining $C_m=C_{\mathrm{tot}}-C_p$ (no electronic term is expected
for this insulating oxide). The resulting $C_m(T)$ shows a broad, non-divergent maximum
($C_m^{\mathrm{max}}\sim 1.6~\mathrm{J\,mol^{-1}\,K^{-1}}$) and then decreases approximately as $T^{-1}$ up to
$\sim0.3$~K, with no additional anomalies between 0.3 and 5~K. Integration of $C_m/T$ reveals that the entropy
release is strongly incomplete in zero field: by \SI{15}{K} we recover
$S_m \simeq 2.59~\mathrm{J\,mol^{-1}\,K^{-1}} \approx 0.45\,R\ln2$, leaving a missing entropy of
$R\ln2-S_m \simeq 3.17~\mathrm{J\,mol^{-1}\,K^{-1}} \approx 0.55\,R\ln2$ within our experimental window down
to \SI{40}{mK} [Fig.~\ref{fig:specific_heat}(d)]. This large entropy deficit, together with the absence of a
transition feature, points to persistent low-energy correlations and/or fluctuating degrees of freedom that
retain a substantial fraction of the ground-state entropy below the base temperature. This interpretation is further strengthened by the absence of any ordering or spin-freezing anomalies down to \SI{50}{mK} in $\chi_{AC}$ investigations of the isostructural analogue NdZnAl$_{11}$O$_{19}$~\cite{cao2025}.

To probe the field evolution of the low-energy spectrum, we measured $C(T)$ in magnetic fields up to \SI{3}{T}
with $H\parallel c$ and extracted $C_m(T)$ [Fig.~\ref{fig:specific_heat}(b)]. With increasing field, the
broad low-$T$ anomaly shifts to higher temperature and becomes more Schottky-like, consistent with Zeeman
splitting of the ground-state Kramers doublet and progressive suppression of correlation effects as the
response becomes increasingly single-ion dominated \cite{Rao2021,Kumar2025,Khatua2023,Kimchi2018}. In this
regime we fit $C_m(T)$ using a two-level Schottky model,
\begin{equation}
C_m(T)=\eta R\left(\frac{\Delta}{T}\right)^2
\frac{e^{-\Delta/T}}{\left(1+e^{-\Delta/T}\right)^2},
\end{equation}
where $\Delta$ is the Zeeman splitting expressed in Kelvin (i.e., $\Delta \equiv \Delta_E/k_B$), and $\eta$
quantifies the fraction of Nd$^{3+}$ ions contributing to the two-level anomaly. The solid curves in
Fig.~\ref{fig:specific_heat}(b) show that this model describes the data well for fields
$0.75$--\SI{3}{T}. In contrast, for $H\lesssim0.75$~T the Schottky description becomes unstable: the fitted
$\eta$ drops sharply and the extracted gap becomes ill-defined, signaling that weak fields do not fully
quench the dominant correlations responsible for the zero-field anomaly (and thus the system is not yet in a
simple two-level regime).

From the reliable Schottky fits ($0.75$--\SI{3}{T}) we obtain $\Delta(H)$ and find a linear field dependence
[inset of Fig.~\ref{fig:specific_heat}(c)].
Fitting $\Delta(H)=\Delta_0+(g_{\mathrm{eff}}\mu_B/k_B)\mu_0H$ over $0.75$--\SI{3}{T} yields
$g_{\mathrm{eff}}=3.86(5)$ with $\Delta_0$ consistent with zero within uncertainty, in agreement with the
single-ion anisotropy inferred independently from Curie--Weiss analysis, Brillouin-function fits,
point-charge crystal-field estimates (Appendix~\ref{app:cefpc}), and EPR (Appendix~\ref{app:epr}).

The fitted $\eta$ remains close to unity at high fields ($\eta\simeq0.82$--0.87 between 1 and \SI{3}{T}),
indicating a modest reduction of the effective Nd$^{3+}$ spectral weight (consistent with a small Nd
deficiency; Appendix~\ref{app:struct}). This is corroborated by the magnetic entropy at \SI{3}{T}: the uncorrected $S_m(T)$ approaches
$S_m\simeq 4.98~\mathrm{J\,mol^{-1}\,K^{-1}}$ by \SI{15}{K}, i.e.\ it falls short of $R\ln2$ by
$\sim0.78~\mathrm{J\,mol^{-1}\,K^{-1}}$, matching the deficit expected from $\eta<1$
[Fig.~\ref{fig:specific_heat}(d)]. After accounting for this missing spectral weight (gray shaded region), the
\SI{3}{T} entropy is consistent with $R\ln2$ by \SI{15}{K}, supporting the assignment of an effective
$S_{\mathrm{eff}}=\tfrac{1}{2}$ Kramers-doublet ground state. The remaining entropy shortfall in zero field
beyond the Nd-deficiency contribution (magenta shaded region) reflects residual entropy retained below
\SI{40}{mK}, consistent with persistent correlated fluctuations rather than symmetry breaking.

\subsection{Adiabatic demagnetisation refrigerant potential}

To directly assess the suitability of NdMgAl$_{11}$O$_{19}$ for ADR, we performed a self-cooling (demagnetisation) experiment in a Quantum Design PPMS using a pressed disk of powdered sample.

\begin{figure}
   \includegraphics[width=\columnwidth]{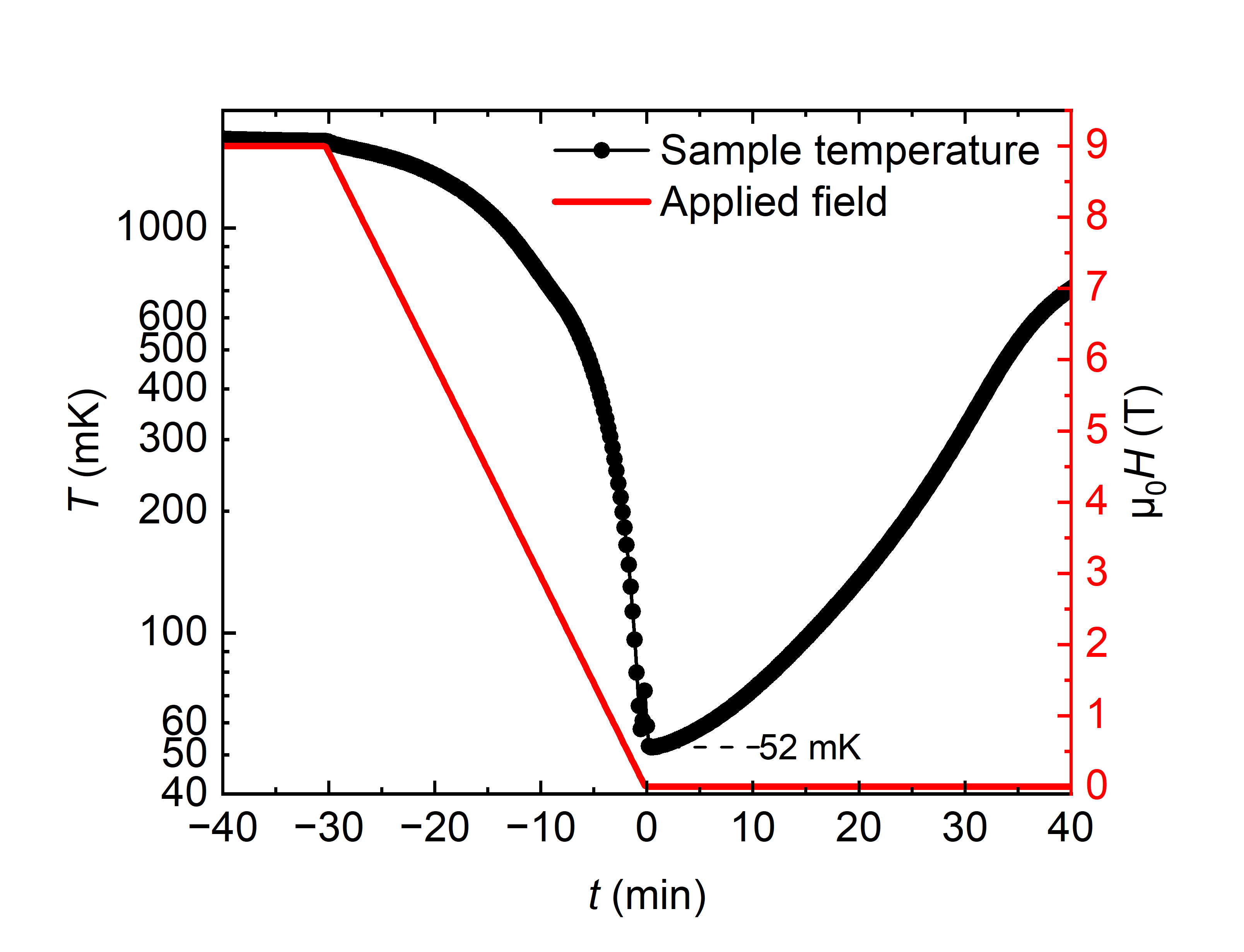}
  \caption{Self-cooling curve for a pressed disc of NdMgAl$_{11}$O$_{19}$ powder during adiabatic demagnetisation from 9 T at a start temperature of 1.8 K.}
  \label{fig:ADR}
\end{figure}

After stabilising the sample at $T_i = 1.8$~K in an applied field of $\mu_0H = 9$~T, the residual helium gas used for thermalisation was absorbed a custom-built in-situ cryopump. Under these quasi-adiabatic conditions, the field was ramped continuously down to zero over 30~minutes, corresponding to an average sweep rate $|d(\mu_0H)/dt| \approx 0.3$~T~min$^{-1}$.
As shown in Fig.~\ref{fig:ADR}, the sample temperature decreased significantly during the field ramp, reaching a minimum temperature of $T_f = 52$~mK at $\mu_0H = 0$~T.

This corresponds to a self-cooling factor of $T_i/T_f \approx 35$, demonstrating that substantial magnetic entropy remains available for refrigeration well below 1~K.
The pronounced cooling is consistent with the low-temperature heat-capacity anomaly and the incomplete release of $R\ln 2$ entropy in zero field, indicating a dense manifold of low-energy magnetic degrees of freedom that can be efficiently polarised at high field and exploited for ADR upon demagnetisation. The non-ideal adiabaticity of the setup then results in a slow-warming curve as heat is leaked back to the sample from the measurement puck's thermal bath.

\section{Discussion}

The combination of SCXRD (Appendix~\ref{app:struct}), magnetization, and specific-heat measurements on
NdMgAl$_{11}$O$_{19}$, together with point-charge CEF calculations (Appendix~\ref{app:cefpc}), reveals a
triangular-lattice rare-earth magnet in which geometric frustration, weak exchange interactions, and strong
single-ion anisotropy cooperatively suppress conventional magnetic order down to at least
\SI{40}{\milli\kelvin}. Magnetization data exhibit pronounced anisotropy, and low-temperature Curie--Weiss fits
(valid when only the ground-state Kramers doublet is populated) yield
$\theta_{\mathrm{CW}}^{ab}\approx -\SI{0.87}{\kelvin}$ and $\theta_{\mathrm{CW}}^{c}\approx -\SI{0.54}{\kelvin}$,
indicating weak antiferromagnetic coupling that is stronger in the $ab$ plane. Brillouin-function fits reproduce
the field-dependent magnetization with high fidelity for both orientations, confirming that the system behaves
as a collection of well-defined, only weakly interacting Kramers pseudospins
\cite{Rao2021,Khatua2022,Khatua2023}.

Despite its paramagnetic behavior above \SI{2}{\kelvin}, the specific heat displays a broad anomaly centered at
\SI{81}{\milli\kelvin}, signaling the development of short-range correlations among Nd$^{3+}$ moments. Crucially,
no sharp $\lambda$-type feature or other indication of long-range magnetic order appears down to
\SI{40}{\milli\kelvin}, implying that any ordering transition lies below our experimental base temperature or is
entirely suppressed by frustration and anisotropy.

Using the direction-dependent Weiss temperatures and noting that no ordering occurs down to $T_N<\SI{40}{\milli\kelvin}$,
we obtain frustration indices
$f_{ab}=|\theta_{\mathrm{CW}}^{ab}|/T_N>0.87/0.04\approx 22$ and
$f_{c}=|\theta_{\mathrm{CW}}^{c}|/T_N>0.54/0.04\approx 13$, both above 10 and underscoring strong magnetic
frustration in NdMgAl$_{11}$O$_{19}$.

In zero applied field, only a portion of the spin-$\tfrac{1}{2}$ entropy is released below \SI{1}{\kelvin},
indicating that a substantial fraction of the ground-state entropy remains locked into low-energy degrees of
freedom. This is consistent with persistent spin dynamics observed in the isostructural analogue
NdZnAl$_{11}$O$_{19}$ using $\mu$SR down to \SI{0.5}{\kelvin}~\cite{cao2025}. Between \SI{1}{\kelvin} and
\SI{6}{\kelvin} the entropy shows no further evolution, consistent with a frustration-stabilized manifold without
any intervening ordering or freezing transition. Only at higher temperatures does the entropy begin to increase
again due to thermal population of the first excited crystal-field doublet (Appendix~\ref{app:cefpc}).

When a \SI{3}{\tesla} magnetic field is applied along the $c$ axis, a Zeeman gap of
$\Delta \approx \SI{0.67}{\milli\electronvolt}$ opens, producing a Schottky anomaly near \SI{4}{\kelvin}. This
field lifts the ground-state degeneracy and restores nearly the full $R\ln2$ entropy by \SI{15}{\kelvin}, with an
additional entropy release compared to zero field. The field-induced recovery of entropy confirms that the
``missing'' entropy at zero field resides in a macroscopically degenerate, frustration-driven manifold rather
than in inaccessible CEF levels, in close analogy to observations in Ba$_3$Yb(BO$_3$)$_3$~\cite{Bag2021}.

The 9.2(4)\% Nd-site vacancies ($p=0.908$; Appendix~\ref{app:struct}) will further affect the magnetic
properties. Since the occupation level is well above the two-dimensional site percolation threshold
($p_c=0.5$), the triangular lattice remains fully connected. In the hypothetical undiluted limit ($p=1$),
$|\theta_{\mathrm{CW}}|$ would increase by approximately 10\%, indicating only a minor renormalization effect due
to the dilution observed from diffraction.

Classical Monte Carlo simulations of the site-diluted triangular-lattice Ising antiferromagnet have long shown
that nonmagnetic vacancies liberate nearly free $S=\tfrac{1}{2}$ ``orphan'' spins: early studies by Grest and
Gabl demonstrated defect-induced spin-glass freezing in this model \cite{Grest1979}, and Creswick identified a
corresponding depletion of zero--net-field sites in the local-field distribution \cite{Creswick1985}. Building
on these, Moessner and Berlinsky quantified that each Henley-type defect triad (three isolated vacancies)
produces one orphan moment, leading to the rule of thumb of roughly one orphan per three vacancies
\cite{Moessner1999}. A simple geometric estimate then gives
\[
n_{\mathrm{orphan}} \approx \frac{1}{3}(1-p)=0.031
\quad\text{per Nd site},
\]
and the corresponding maximum entropy is
\[
S_{\mathrm{orphan}} = n_{\mathrm{orphan}}\,R\ln 2 \approx 0.18\;\mathrm{J\,mol^{-1}\,K^{-1}},
\]
which remains far below the missing $R\ln2$ entropy we observe up to \SI{15}{\kelvin} in zero field. Moreover,
in a quantum Kramers-ion system, zero-point fluctuations and transverse exchange will further suppress the orphan
contribution, underscoring that degenerate correlated states resulting from geometric frustration, not
dilution-induced orphan spins, dominantly retain the residual entropy.

In mean-field theory for a triangular lattice ($z=6$) with an effective pseudospin
$S_{\mathrm{eff}}=\tfrac{1}{2}$, the Curie--Weiss temperature is related to the nearest-neighbour exchange scale via
\begin{equation}
  \theta_{\mathrm{CW}}
  = \frac{z\,J_{\mathrm{nn}}\,S_{\mathrm{eff}}(S_{\mathrm{eff}}+1)}{3k_B}.
  \label{eq:thetaCW_MF}
\end{equation}
Rearranging Eq.~\eqref{eq:thetaCW_MF} yields $J_{xy}=-0.57(1)\,\mathrm{K}$ and $J_{z}=-0.36(1)\,\mathrm{K}$.
Throughout, we adopt the Hamiltonian convention
$\mathcal{H}=-\sum_{\langle ij\rangle}\sum_{\alpha=x,y,z} J_{\alpha}\,S_i^{\alpha}S_j^{\alpha}$, such that
antiferromagnetic coupling corresponds to $J_{\alpha}<0$. We then define the isotropic average
\[
  J=\frac{J_{z}+2J_{xy}}{3}=-0.50(1)\,\mathrm{K},
\]
and the exchange-anisotropy parameter
\[
  D=J_{z}-J=-0.14(1)\,\mathrm{K},
\]
so that $D<0$ indicates easy-plane exchange anisotropy and $|D/J|\approx 0.28$.

Because $D<0$, the system is pushed toward an effective two-dimensional XY limit, consistent with an easy-plane
XXZ pseudospin model, which in the absence of explicit symmetry breaking may exhibit a
Berezinskii--Kosterlitz--Thouless (BKT) vortex-unbinding crossover rather than a conventional phase transition
\cite{Berezinskii1971,Kosterlitz1973,Zhang2025}. Alternatively, the partial recovery of magnetic entropy and the
persistent low-$T$ spin fluctuations could instead reflect a quantum-disordered ground state or a short-range--ordered
XXZ antiferromagnet strongly influenced by geometric frustration. Crucially, the exchange constant $J_{\rm ex}$
extracted from a mean-field Curie--Weiss analysis represents only a bulk-averaged scale; it can underestimate
competing longer-range couplings and neglect quantum-renormalization effects that are endemic to frustrated lattices.

The pronounced self-cooling observed under quasi-adiabatic demagnetisation (Fig.~\ref{fig:ADR}) places
NdMgAl$_{11}$O$_{19}$ within a broader class of low-temperature magnetocaloric materials in which frustration and/or
disorder suppress long-range order and retain a sizable entropy reservoir to deeply sub-Kelvin temperatures
\cite{Treu2025,Dwivedi2025}. A key limitation of the present compound, however, is its comparatively low rare-earth
moment density: even if the full Kramers-doublet entropy $R\ln2$ per Nd is available, the maximum entropy density is
\[
  s_{\mathrm{max}}=\frac{Z R\ln 2}{N_A V_{\mathrm{cell}}}\simeq 32.5\,\mathrm{mJ\,cm^{-3}\,K^{-1}},
\]
using $Z=2$ and the refined hexagonal unit cell (Appendix~\ref{app:struct}). This is comparable to historic
hydrated-salt refrigerants \cite{Tokiwa2021}, but remains modest compared with dense Gd$^{3+}$-based refrigerants,
where the spin-only $R\ln(8)$ entropy per ion combined with a high concentration of magnetic ions yields
substantially larger volumetric entropy changes and cooling power; recent frustrated Gd-rich oxides in particular
demonstrate state-of-the-art low-temperature performance \cite{Zhang2026,Yang2026}, alongside established benchmarks
such as rare-earth garnets \cite{Kleinhans2023} and other chemically robust Gd-based lattices
\cite{Delacotte2022,Shanta2025}. Nevertheless, the strong magnetic anisotropy of NdMgAl$_{11}$O$_{19}$, coupled with
the availability of reasonably sized single crystals, makes this system a promising candidate for exploring
rotational magnetocaloric effects.

A definitive determination of the true ground state will require local probes. Inelastic neutron scattering can map
out the full spin-excitation spectrum; $\mu$SR or NMR can resolve slow dynamics and possible critical slowing down;
and torque magnetometry (or neutron spin-echo) can directly access the spin stiffness and test for the hallmark
universal jump expected if the system follows a BKT-like crossover. Taken together, our thermodynamic and
magnetization data place NdMgAl$_{11}$O$_{19}$ close to a 2D-XY regime in which a BKT-type vortex-binding crossover
might occur; however, only a combination of such local-probe measurements and theory constrained by the
experimentally determined exchange parameters can unambiguously distinguish BKT physics from alternative
quantum-correlated or quantum-disordered scenarios, such as quantum spin liquid, valence bond solid, or random
singlet ground states.

In the discussion, we considered the possible effects of Nd deficiency on the magnetic and thermodynamic response;
however, for clarity and simplicity, heat-capacity and magnetic-property data are plotted and analyzed per mol
formula unit, i.e., neglecting the Nd deficiency.

\section{Conclusion}

Our combined SCXRD, magnetization, specific-heat measurements on single-crystalline NdMgAl$_{11}$O$_{19}$ supported by point charge calculation reveal a triangular-lattice system composed of well-isolated Kramers doublets with strong axial anisotropy. The extracted $\theta_{\rm CW} \approx -0.54$\,K indicates the presence of antiferromagnetic interactions, yet no evidence of long-range magnetic order or spin-glass freezing is observed down to our base temperature of 40\,mK. This robust suppression of ordering, despite magnetic exchange and measurable structural disorder, points to strong geometric frustration amplified by the system’s quasi-two-dimensionality and $\sim$9.2\% Nd-site dilution.

The thermal anomaly centered at $\sim$81\,mK, combined with the large residual entropy and its full recovery under applied field, further confirms the presence of a frustrated manifold that allows self-cooling down to 52 mK on demagnetisation from a field of 9 T . The easy-plane XXZ exchange regime inferred from CW analysis and $g$-tensor anisotropy suggests the possibility of a BKT type crossover. However, the proximity to a quantum-disordered state with persistent spin dynamics cannot be excluded, as has been suggested in the isostructural analogue NdZnAl$_{11}$O$_{19}$~\cite{cao2025}. In this context, NdMgAl$_{11}$O$_{19}$ emerges as a promising candidate for a triangular-lattice quantum spin liquid, readily grown as a single crystal. Further experimental investigations, such as low-temperature inelastic neutron scattering and muon spin relaxation will be essential to clarify the nature of its ground state.

\section*{Authors' Contributions}

S.~Kumar prepared the samples, verified crystal quality, reproduced the magnetization and specific-heat datasets, performed point-charge crystal-field calculations, carried out data analysis, and drafted the manuscript. 
G.~Bastien initiated the project, performed the initial magnetization and specific-heat measurements, and supervised parts of the analysis.
J.~Prokleška performed dilution refrigerator measurements of specific heat.
M.~Kempiński and W.~Kempiński conducted the EPR measurements. 
K.~Załęski reproduced magnetization measurements on additional single crystals. 
A.~Kancko and C.~Correa performed the single-crystal X-ray diffraction measurements and structural analysis. 
T.~Treu and P.~Gegenwart carried out the ADR performance tests. 
M.~Śliwińska-Bartkowiak supervised and coordinated all experimental work performed in Poznań. 
R.~H.~Colman supervised the crystal growth, data analysis and manuscript preparation.

\begin{acknowledgments}
We acknowledge funding from Charles University in Prague within the Primus research program with grant number PRIMUS/22/SCI/016, and the Grant Agency of Univerzita Karlova (grant number 438425). The work was also supported by the Ministry of Education, Youth and Sports of the Czech Republic through program INTER-EXCELLENCE II INTER-ACTION (LUABA24056) and the Czech Science Foundation (Project No.26-23051S). Work at the University of Augsburg was supported by the Deutsche Forschungsgemeinschaft (DFG, German ResearchFoundation), Grant No. 514162746 (GE 1640/11–1) and the Bavarian-Czech Academic Agency BTHA, Grant No. JC-2024-15 (BaCQuERel). Crystal growth, structural analysis, and magnetic properties measurements were carried out in the MGML (\url{http://mgml.eu/}), supported within the Czech Research Infrastructures program (project no. LM2023065).
\end{acknowledgments}

\begin{appendices}

\appendix

\section{Structural Analysis}\label{app:struct}

Single crystal X-ray diffraction was performed at 95 K on a Rigaku SuperNova diffractometer equipped with an Atlas S2 CCD detector, using mirror-collimated Mo K$\alpha$ ($\lambda$ = 0.71073 \AA) radiation from a micro-focus sealed tube. Diffraction data were integrated using CrysAlis Pro \cite{crysalispro}, combining an empirical absorption correction using spherical harmonics \cite{clark1995}, and a numerical absorption correction based on Gaussian integration over a multifaceted crystal model, implemented in the SCALE3 ABSPACK scaling algorithm. The structure was solved by charge flipping using the program Superflip \cite{palatinus2007} and refined by full-matrix least-squares on $F^2$ in Jana2020 \cite{petricek2023}. Structural graphics were created using Jana2020 and Vesta \cite{momma2008}.

The crystal structure of NdMgAl$_{11}$O$_{19}$ obtained by SCXRD at \SI{95}{\kelvin} is hexagonal, space group \(P6_3/mmc\) (\#194), \(Z = 2\), with unit cell parameters \(a = \SI{5.5771(3)}{\angstrom}\) and \(c = \SI{21.8542(12) }{\angstrom}\), consistent with the expected magnetoplumbite structure reported in previous $RE$MgAl$_{11}$O$_{19}$ family members ($RE$ = Ce, Pr) \citep{gao2024,kumar2024}. Due to the nearly identical electronic configurations of Mg$^{2+}$ and Al$^{3+}$, and therefore very similar X-ray atomic scattering factors, it is very difficult to distinguish the site of Mg$^{2+}$ ions via SCXRD. Therefore, an initial model of NdAl$_{12}$O$_{19}$ was assumed, leading to several negative atomic displacement parameters (ADPs). Inspection of the Fourier difference maps indicates a residual density of 1.66 $e$\AA$^{-3}$ at 0.83 \AA {} from Nd(1). Previous SCXRD studies of related $RE$MgAl$_{11}$O$_{19}$  members (RE = Ce, Pr) suggested an additional improvement of the model by adding a RE(2) atom to the $6h$ site at a distance of $\sim$0.85 \AA {} from RE(1) located on the $2d$ site \cite{gao2024,kumar2024}. Inclusion of this positionally disordered (off-centred) Nd(2) ion results in a non-negligible improvement of the model, with positively defined ADPs, refined occupancies occ(Nd1) = 0.852(4) and occ(Nd2) = 0.0188(9), and agreement factors GOF(obs) =  1.16\% and $R$(obs) = 2.04\%. Previous neutron diffraction studies in the isostructural analogue CeMgAl$_{11}$O$_{19}$ \cite{gao2024} suggest that Mg$^{2+}$  is shared with Al$^{3+}$ on the Al(4) site. Splitting the Al(4) site between Mg and Al with fixed 0.5/0.5 occupancies, with harmonic ADPs and coordinates constrained to be equal, practically did not change the quality of the refinement - GOF(obs) = 1.15\% and $R$(obs) = 2.01\%. In the magnetoplumbite structure, the Al(5) ion is often distributed between two off-centered positions in the oxygen bipyramid, with the $z$-coordinate refined to $z$(Al5) = 0.2430(13), resulting in the off-centering distance $\delta = 2(0.25 - z$(Al5))$c$ = 0.31 \AA.  

This solution gives the final formula Nd$_{0.908}$MgAl$_{11}$O$_{19}$, with 85.2(4)\% Nd occupying the high-symmetry $2d$ site and 1.88(9)\% Nd occupying the off-centered $6h$ site. The final composition suggests a $\sim$9.2(4)\% total Nd$^{3+}$ deficiency. Further structural details are available 
in the deposited CIF file at the Cambridge Crystallographic Data Centre (CCDC) under 
accession code 2447341 (\url{https://www.ccdc.cam.ac.uk}).

\begin{sidewaystable}[]
\caption{Single crystal X-ray diffraction experimental results at $T$ = 95 K, and structural refinement of NdMgAl$_{11}$O$_{19}$}
\renewcommand{\arraystretch}{1.1}
\setlength{\tabcolsep}{4pt}
\begin{tabular}{cccccccccccccccccccccccccc}
\hline
\multicolumn{4}{c}{\begin{tabular}[c]{@{}c@{}} Chemical composition \\ Nd$_{0.908}$MgAl$_{11}$O$_{19}$ \end{tabular}} &
\multicolumn{5}{c}{\begin{tabular}[c]{@{}c@{}}Crystal system, space group\\
hexagonal, $P6_3/mmc$ (\#194)\\
$Z$ = 2\end{tabular}} &
\multicolumn{6}{c}{\begin{tabular}[c]{@{}c@{}}$a$ = 5.5771(3) $\mathring{\mathrm{A}}$\\
$c$ = 21.8542(12) $\mathring{\mathrm{A}}$\\
$V$ = 588.67(6) $\mathring{\mathrm{A}}^3$\end{tabular}} &
\multicolumn{7}{c}{\begin{tabular}[c]{@{}c@{}}Crystal size\\
45$\times$17$\times$11 $\mu$m$^3$\end{tabular}} &
\multicolumn{4}{c}{\begin{tabular}[c]{@{}c@{}}Density (calculated)\\
4.2656 g\,cm$^{-3}$\end{tabular}} \\
\hline

\multicolumn{3}{c}{\begin{tabular}[c]{@{}c@{}}X-ray tube: Mo K$\alpha$\\
($\lambda$ = 0.71073 $\mathring{\mathrm{A}}$)\\
$T$ = 95 K\end{tabular}} &
\multicolumn{4}{c}{\begin{tabular}[c]{@{}c@{}}No. of reflections\\
collected/unique/used:\\
8607/358/328\\
$R_{\rm int}=4.85 \%$ \end{tabular}} &
\multicolumn{4}{c}{\begin{tabular}[c]{@{}c@{}}Final indices\\
$R_{obs}$ = 2.01\%\\
$wR_2$ = 5.88\%\\
GOF = 1.15 \\
Parameters: 45 \end{tabular}} &
\multicolumn{4}{c}{\begin{tabular}[c]{@{}c@{}}Index ranges: \\
$-7 \leq h \leq 7$ \\
$-7 \leq k \leq 6$ \\ 
$-28 \leq l \leq 27$ \\
$\theta$ range: \\ 
3.73$^\circ$-29.62$^\circ$ \end{tabular}}&
\multicolumn{7}{c}{\begin{tabular}[c]{@{}c@{}}$\Delta\rho_{max}$, $\Delta\rho_{min}$ ($e\mathring{\mathrm{A}}^{-3}$)\\
0.40, $-$0.75\end{tabular}} &
\multicolumn{4}{c}{\begin{tabular}[c]{@{}c@{}}Absorption\\
$\mu$ = 5.041 mm$^{-1}$\end{tabular}} \\
\hline

\multirow{2}{*}{Atom} & \multirow{2}{*}{Site} & \multirow{2}{*}{Symmetry} &
\multicolumn{3}{c}{\centering $x$} &
\multicolumn{2}{c}{\centering $y$} &
\multicolumn{1}{c}{\centering $z$} &
\multicolumn{2}{c}{\centering Occupancy} &
\multicolumn{15}{c}{$U_{ij}$ ($\times 10^{-4} \mathring{\mathrm{A}}^2$)} \\
\cline{12-26}
 & & & & & & & & &
 & &
\multicolumn{2}{c}{$U_{11}$} & \multicolumn{2}{c}{$U_{22}$} &
\multicolumn{2}{c}{$U_{33}$} &
\multicolumn{3}{c}{$U_{12}$} &
\multicolumn{2}{c}{$U_{13}$} &
\multicolumn{2}{c}{$U_{23}$} &
\multicolumn{2}{c}{$U_{eq}$} \\
\hline

Nd(1) & 2$d$ & $\overline{6}m2$ &
\multicolumn{3}{c}{$\frac{1}{3}$} & \multicolumn{2}{c}{$\frac{2}{3}$} & $\frac{3}{4}$ &
\multicolumn{2}{c}{0.852(4)} &
\multicolumn{2}{c}{68(3)} & \multicolumn{2}{c}{$U_{11}$} &
\multicolumn{2}{c}{32(3)} &
\multicolumn{3}{c}{$\frac{1}{2}U_{11}$} &
\multicolumn{2}{c}{0} & \multicolumn{2}{c}{0} &
\multicolumn{2}{c}{56(2)} \\

Nd(2) & 6$h$ & $m$ &
\multicolumn{3}{c}{0.529(9)} & \multicolumn{2}{c}{0.264(5)} & $\frac{1}{4}$ &
\multicolumn{2}{c}{0.0188(9)} &
\multicolumn{2}{c}{68(3)} & \multicolumn{2}{c}{$U_{11}$} &
\multicolumn{2}{c}{32(3)} &
\multicolumn{3}{c}{$\frac{1}{2}U_{11}$} &
\multicolumn{2}{c}{0} & \multicolumn{2}{c}{0} &
\multicolumn{2}{c}{56(2)} \\

Al(1) & 2$a$ & $\overline{3}m$ &
\multicolumn{3}{c}{0} & \multicolumn{2}{c}{0} & 0.5 &
\multicolumn{2}{c}{1} &
\multicolumn{2}{c}{25(4)} & \multicolumn{2}{c}{$U_{11}$} &
\multicolumn{2}{c}{20(10)} &
\multicolumn{3}{c}{$\frac{1}{2}U_{22}$} &
\multicolumn{2}{c}{0} & \multicolumn{2}{c}{0} &
\multicolumn{2}{c}{24(5)} \\

Al(2) & 4$f$ & $3m$ &
\multicolumn{3}{c}{$\frac{1}{3}$} & \multicolumn{2}{c}{$\frac{2}{3}$} & 0.30991(7) &
\multicolumn{2}{c}{1} &
\multicolumn{2}{c}{28(5)} & \multicolumn{2}{c}{$U_{11}$} &
\multicolumn{2}{c}{30(7)} &
\multicolumn{3}{c}{$\frac{1}{2}U_{11}$} &
\multicolumn{2}{c}{0} & \multicolumn{2}{c}{0} &
\multicolumn{2}{c}{29(4)} \\

Al(3) & 12$k$ & $m$ &
\multicolumn{3}{c}{0.83231(6)} & \multicolumn{2}{c}{0.66462(12)} & 0.39149(4) &
\multicolumn{2}{c}{1} &
\multicolumn{2}{c}{28(4)} & \multicolumn{2}{c}{30(4)} &
\multicolumn{2}{c}{37(5)} &
\multicolumn{3}{c}{$\frac{1}{2}U_{11}$} &
\multicolumn{2}{c}{-12(11)} & \multicolumn{2}{c}{-2(2)} &
\multicolumn{2}{c}{31(3)} \\

Al(4) & 4$f$ & $3m$ &
\multicolumn{3}{c}{$\frac{1}{3}$} & \multicolumn{2}{c}{$\frac{2}{3}$} & 0.47258(7) &
\multicolumn{2}{c}{1/2} &
\multicolumn{2}{c}{15(5)} & \multicolumn{2}{c}{$U_{11}$} &
\multicolumn{2}{c}{21(8)} &
\multicolumn{3}{c}{$\frac{1}{2}U_{11}$} &
\multicolumn{2}{c}{0} & \multicolumn{2}{c}{0} &
\multicolumn{2}{c}{17(4)} \\

Mg(1) & 4$f$ & $3m$ &
\multicolumn{3}{c}{$\frac{1}{3}$} & \multicolumn{2}{c}{$\frac{2}{3}$} & 0.47258(7) &
\multicolumn{2}{c}{1/2} &
\multicolumn{2}{c}{15(5)} & \multicolumn{2}{c}{$U_{11}$} &
\multicolumn{2}{c}{21(8)} &
\multicolumn{3}{c}{$\frac{1}{2}U_{11}$} &
\multicolumn{2}{c}{0} & \multicolumn{2}{c}{0} &
\multicolumn{2}{c}{17(4)} \\

Al(5) & 4$e$ & $3m$ &
\multicolumn{3}{c}{0} & \multicolumn{2}{c}{0} & 0.2430(13) &
\multicolumn{2}{c}{1/2} &
\multicolumn{2}{c}{45(6)} & \multicolumn{2}{c}{$U_{11}$} &
\multicolumn{2}{c}{17(11)} &
\multicolumn{3}{c}{$\frac{1}{2}U_{11}$} &
\multicolumn{2}{c}{0} & \multicolumn{2}{c}{0} &
\multicolumn{2}{c}{90(40)} \\

O(1) & 6$h$ & $mm2$ &
\multicolumn{3}{c}{0.6365(4)} & \multicolumn{2}{c}{0.8182(2)} & $\frac{1}{4}$ &
\multicolumn{2}{c}{1} &
\multicolumn{2}{c}{50(12)} & \multicolumn{2}{c}{92(10)} &
\multicolumn{2}{c}{56(15)} &
\multicolumn{3}{c}{$\frac{1}{2}U_{11}$} &
\multicolumn{2}{c}{0} & \multicolumn{2}{c}{0} &
\multicolumn{2}{c}{71(9)} \\

O(2) & 4$f$ & $3m$ &
\multicolumn{3}{c}{$\frac{1}{3}$} & \multicolumn{2}{c}{$\frac{2}{3}$} & 0.44220(17) &
\multicolumn{2}{c}{1} &
\multicolumn{2}{c}{27(9)} & \multicolumn{2}{c}{27(9)} &
\multicolumn{2}{c}{81(18)} &
\multicolumn{3}{c}{$\frac{1}{2}U_{22}$} &
\multicolumn{2}{c}{0} & \multicolumn{2}{c}{0} &
\multicolumn{2}{c}{45(9)} \\

O(3) & 4$e$ & $3m$ &
\multicolumn{3}{c}{0} & \multicolumn{2}{c}{0} & 0.34888(14) &
\multicolumn{2}{c}{1} &
\multicolumn{2}{c}{29(8)} & \multicolumn{2}{c}{$U_{11}$} &
\multicolumn{2}{c}{37(17)} &
\multicolumn{3}{c}{$\frac{1}{2}U_{11}$} &
\multicolumn{2}{c}{0} & \multicolumn{2}{c}{0} &
\multicolumn{2}{c}{32(8)} \\

O(4) & 12$k$ & $m$ &
\multicolumn{3}{c}{0.6950(3)} & \multicolumn{2}{c}{0.84748(16)} & 0.44653(10) &
\multicolumn{2}{c}{1} &
\multicolumn{2}{c}{67(8)} & \multicolumn{2}{c}{43(7)} &
\multicolumn{2}{c}{45(10)} &
\multicolumn{3}{c}{$\frac{1}{2}U_{22}$} &
\multicolumn{2}{c}{-12(6)} & \multicolumn{2}{c}{-6(3)} &
\multicolumn{2}{c}{49(6)} \\

O(5) & 12$k$ & $m$ &
\multicolumn{3}{c}{0.50539(16)} & \multicolumn{2}{c}{0.49461(16)} & 0.34855(8) &
\multicolumn{2}{c}{1} &
\multicolumn{2}{c}{21(6)} & \multicolumn{2}{c}{$U_{11}$} &
\multicolumn{2}{c}{77(11)} &
\multicolumn{3}{c}{6(6)} &
\multicolumn{2}{c}{-5(3)} & \multicolumn{2}{c}{5(3)} &
\multicolumn{2}{c}{42(6)} \\

\hline
\end{tabular}
\end{sidewaystable}

\FloatBarrier
\clearpage

\section{Magnetic susceptibility}\label{app:susceptibility}

\begin{figure}[t]
\centering
\includegraphics[width=0.48\textwidth]{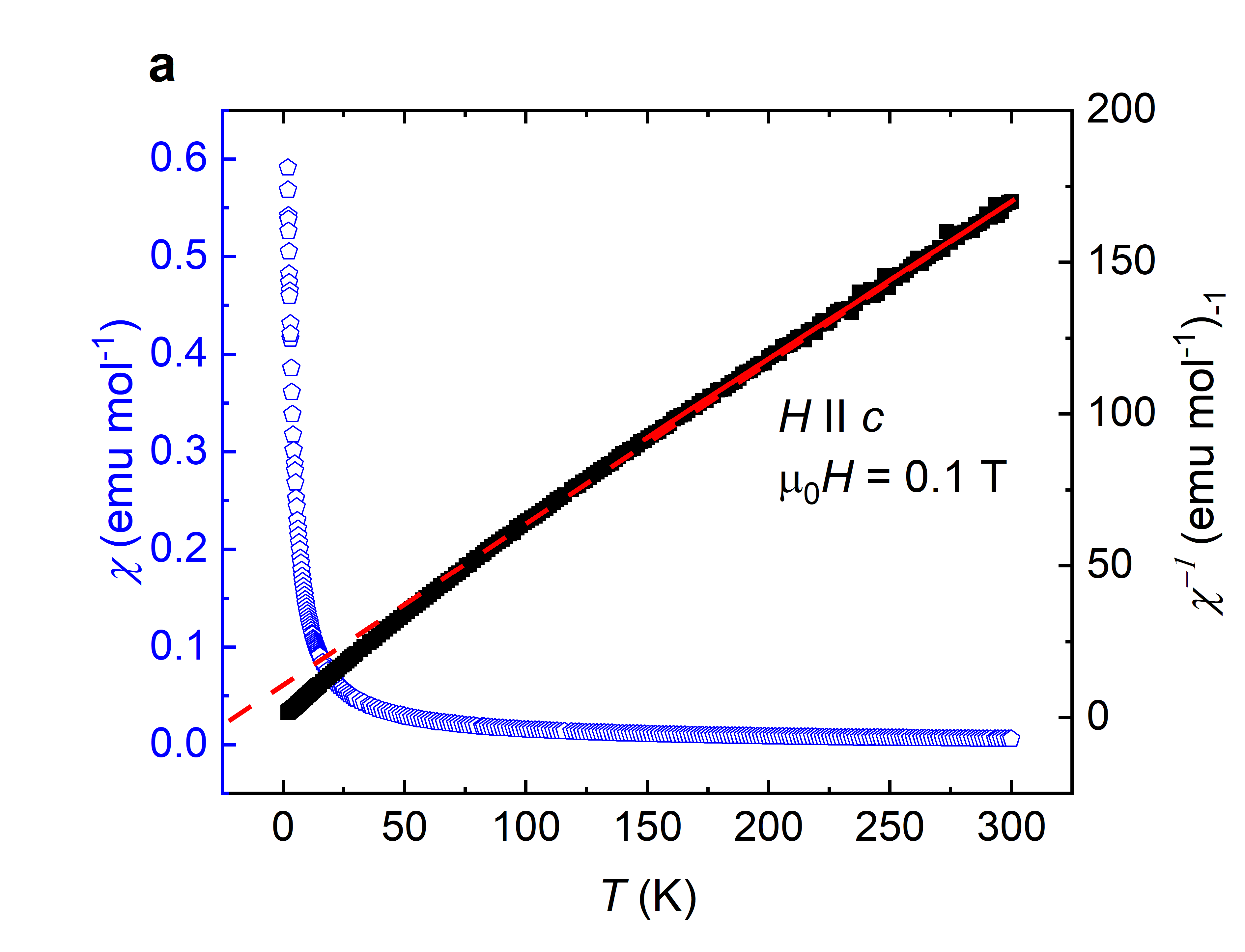}\hfill
\includegraphics[width=0.48\textwidth]{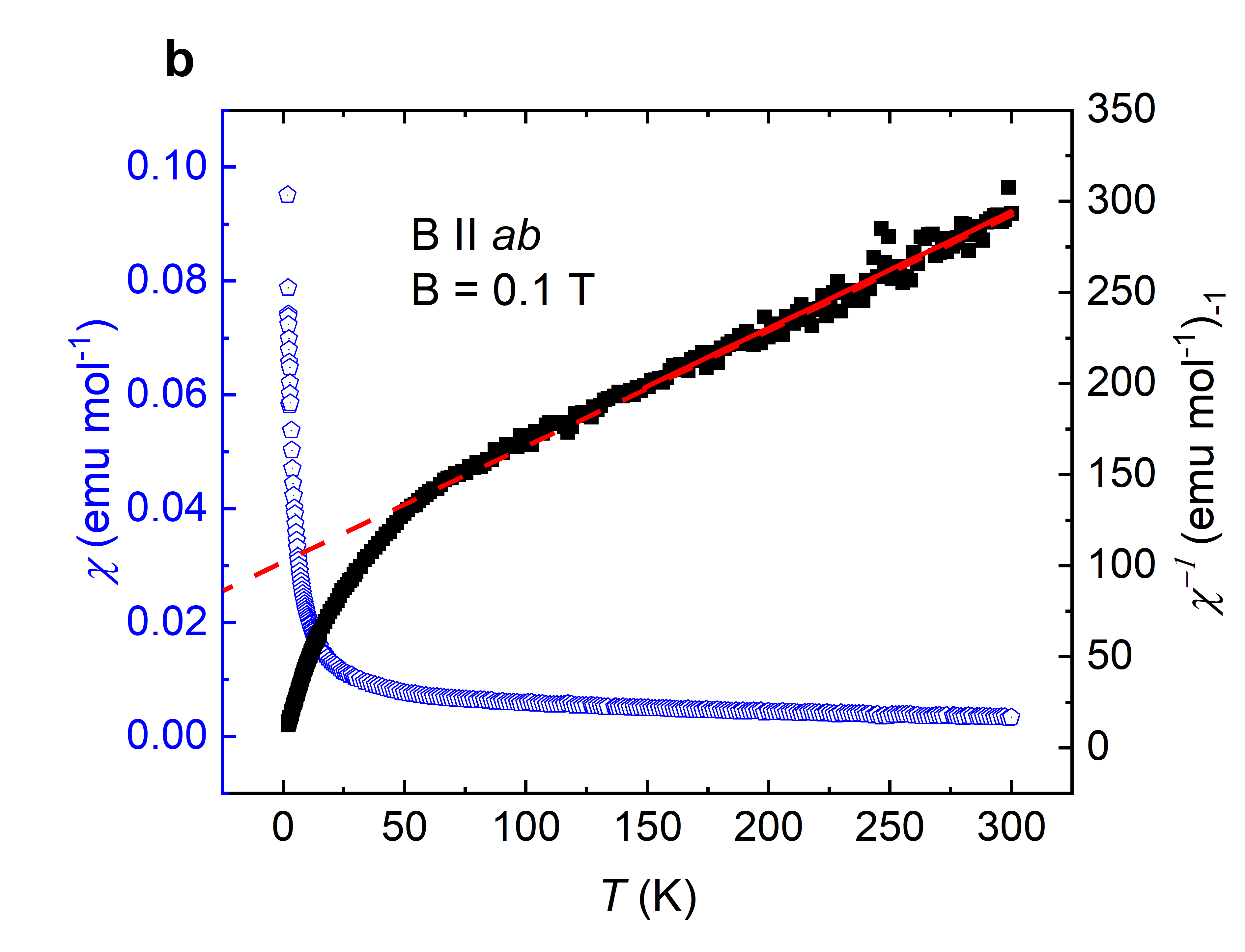}
\caption{
(a) Magnetic susceptibility $\chi(T)$ and inverse susceptibility $\chi^{-1}(T)$ of NdMgAl$_{11}$O$_{19}$
measured with $H \parallel c$. Solid red lines show the corresponding high-temperature Curie--Weiss fits,
while dashed lines highlight the deviation from the linear Curie--Weiss behavior upon heating due to CEF
excitations.
(b) Same as panel (a), but for $H \parallel ab$. Solid red lines indicate the high-temperature Curie--Weiss
fits and dashed lines mark the deviation from linear behavior at elevated temperatures.
}
\label{figS1}
\end{figure}

As discussed in the main text, the inverse susceptibility $\chi^{-1}(T)$ shows a clear deviation from linear Curie--Weiss behavior upon heating(Fig\ref{figS1}). This nonlinearity reflects the
increasing importance of thermally activated crystal-electric-field (CEF) excitations (and associated
Van Vleck--type contributions), which progressively invalidate a single-parameter Curie--Weiss
description outside the strict high-$T$ limit. To quantify this regime, we performed high-temperature
Curie--Weiss fits of $\chi^{-1}(T)$ for both field orientations; the extracted parameters are summarized
in Table~\ref{tab:CW_highT}.

\begin{table}
\centering
\caption{High-temperature Curie--Weiss fit parameters for both crystallographic directions.}
\label{tab:CW_highT}
\begin{tabular}{lcccc}
\hline
Direction & $C$ (emu\,K/mol) & $\theta_{\mathrm{CW}}$ (K) & $\mu_{\mathrm{eff}}$ ($\mu_B$) & $g_J$ ($J=9/2$) \\
\hline
$H \parallel c$ & 1.93(8)  & $-28.3(0.9)$   & 3.93(1) & 0.79(2) \\
$H \parallel ab$     & 1.56(2) & $-160.8(5.9)$  & 3.54(4) & 0.71(7) \\
\hline
\end{tabular}
\end{table}

The extracted $\mu_{\mathrm{eff}}$ and $g_J$ are in reasonable agreement with the Nd$^{3+}$ free-ion values
($\mu_{\mathrm{eff}}^{\mathrm{free}}\approx 3.62\,\mu_B$, $g_J^{\mathrm{free}}=8/11\approx 0.727$), confirming that the
high-temperature response is dominated by localized Nd$^{3+}$ moments. However, the large and strongly
anisotropic Curie--Weiss temperatures obtained from these fits cannot be used to define the low-temperature
interaction scale; in the presence of CEF excitations, the fitted $\theta_{\mathrm{CW}}$ becomes an effective
parameter that absorbs excited-level and background contributions rather than reflecting the microscopic
exchange energy governing the low-$T$ regime.

\section{Point-charge crystal electric field calculation with intermediate coupling}\label{app:cefpc}

To estimate the CEF level scheme and single-ion magnetic anisotropy of NdMgAl$_{11}$O$_{19}$, we performed
point-charge CEF calculations using \textsc{PyCrystalField} \cite{PyCrystalField} for the majority Nd$^{3+}$
site at the high-symmetry Wyckoff position $2d$ with fractional coordinates $[1/3,\,2/3,\,3/4]$.
The full first oxygen coordination shell was included (12 O ligands within 3~\AA), and in the absence of
charges in the CIF the ligands were assumed to be O$^{2-}$ with a central-ion charge Nd$^{3+}$ (standard
point-charge approximation).

In rare-earth ions, crystal-field splitting and the magnetic response are not determined by a rigid isolated
$J$ multiplet: admixture of excited $J$ manifolds (intermediate coupling, or inter-multiplet $J$-mixing)
renormalizes both the CEF scheme and the effective magnetic response. We therefore included a finite
spin--orbit (LS) coupling constant in the Hamiltonian,
\begin{equation}
  \mathcal{H}=\mathcal{H}_{\rm CEF}+\mathcal{H}_{\rm SO},\qquad
  \mathcal{H}_{\rm SO}=\zeta\,\mathbf{L}\cdot\mathbf{S}.
\end{equation}

To document the scale of temperature-independent contributions in the bulk response, the susceptibility was
fitted to the Curie--Weiss form
\begin{equation}
  \chi(T)=\frac{C}{T-\theta_{\rm CW}}+\chi_0,
\end{equation}
yielding $\theta_{\rm CW}=-0.56(9)$~K (consistent with the low-temperature Curie--Weiss analysis in the main text)
and $\chi_0^{c}=4.96\times10^{-4}$~emu/mol for $H\parallel c$. We emphasize that this extracted $\chi_0$ should not
be interpreted as a pure Van~Vleck (inter-multiplet) term: its magnitude is very small (of order $10^{-4}$~emu/mol),
comparable to expected temperature-independent backgrounds such as the core diamagnetism of the full compound and
other field-independent contributions. Consequently, using $\chi_0$ to quantitatively infer an intermediate-coupling
scale (or $\zeta$) is not robust.

Instead, we adopt a representative free-ion (spectroscopic) scale for Nd$^{3+}$.
For the Nd$^{3+}$ ground term ${}^{4}I$ ($L=6$, $S=3/2$), the free-ion separation between the
${}^{4}I_{9/2}$ ground level and the first excited multiplet ${}^{4}I_{11/2}$ is
$\Delta_{11/2-9/2}=1897.07$~cm$^{-1}$, corresponding to $\Delta_{11/2-9/2}\simeq 235.2$~meV \cite{Wyart2006_Nd3_FreeIon_Levels}.
Within the Russell--Saunders expression \cite{Wagner1972_TheoryOfMagnetism}
$E_J=\frac{\zeta}{2}\!\left[J(J+1)-L(L+1)-S(S+1)\right]$,
the $J=\tfrac{9}{2}\rightarrow\tfrac{11}{2}$ splitting is
\begin{equation}
  \Delta_{11/2-9/2}=E_{11/2}-E_{9/2}=\frac{11}{2}\,\zeta,
\end{equation}
so that
\begin{equation}
  \zeta_{\rm free}=\frac{2}{11}\Delta_{11/2-9/2}\simeq 42.77~\mathrm{meV}.
\end{equation}
We therefore use \texttt{LS\_Coupling}$=42.77$~meV in the intermediate-coupling \textsc{PyCrystalField} calculation.

The CEF Hamiltonian was constructed in Stevens-operator form,
\begin{equation}
  \mathcal{H}_{\rm CEF}=\sum_{n,m}B_n^m\,O_n^m(\mathbf{J}),
\end{equation}
and for the site symmetry detected in our coordinate frame the point-charge model yields only four nonzero
Stevens parameters, giving
\begin{equation}
  \mathcal{H}_{\rm CEF}
  = B_2^{0}O_2^{0}+B_4^{0}O_4^{0}+B_6^{0}O_6^{0}+B_6^{6}O_6^{6},
\end{equation}
with
\begin{align}
B_2^{0} &= -1.6224\times10^{-1}, \nonumber\\
B_4^{0} &= -5.5589\times10^{-4}, \nonumber\\
B_6^{0} &= +4.7717\times10^{-5}, \nonumber\\
B_6^{6} &= +3.5008\times10^{-4}.
\end{align}

Diagonalization with \texttt{LS\_Coupling}$=42.77$~meV produces the expected five Kramers doublets of the
Nd$^{3+}$ $J=9/2$ ground multiplet below 50~meV, with (unique) excitation energies
\[
  E_{\rm CEF}=0,\;2.17,\;9.39,\;21.50,\;24.59~\mathrm{meV},
\]
while the first levels associated with the excited $J$-manifold appear only at much higher energies
(first states near $\sim 237$~meV in the present point-charge model). The ground Kramers doublet exhibits a
strongly anisotropic $g$ tensor with principal values
\[
  g_{ab}\simeq 1.35,\qquad g_c\simeq 4.45,
\]
consistent with a dominant easy-axis response along the crystallographic $c$ axis.

\smallskip

\noindent\textit{Comparison to a single-multiplet calculation (no intermediate coupling).}

For completeness, we also repeated the point-charge calculation without LS coupling (i.e., restricting the
Hilbert space to the isolated $J=9/2$ multiplet). In this case \textsc{PyCrystalField} returns additional symmetry-allowed
terms with odd $m$ and a slightly different set of Stevens parameters,
\begin{align}
B_2^{0} &= -2.5811\times10^{-1}, \nonumber\\
B_4^{0} &= -1.3217\times10^{-3}, \nonumber\\
B_4^{3} &=  1.737\times10^{-5}, \nonumber\\
B_6^{0} &=  1.6167\times10^{-4}, \nonumber\\
B_6^{3} &=  5.005\times10^{-7}, \nonumber\\
B_6^{6} &=  1.1861\times10^{-3}.
\end{align}

The corresponding (unique) CEF excitation energies within the $J=9/2$ manifold are
\[
  E_{\rm CEF}^{(J\text{-only})}=0,\;1.53,\;8.86,\;21.20,\;24.34~\mathrm{meV},
\]
i.e., broadly similar level spacings but with a systematic renormalization relative to the intermediate-coupling result.
This comparison highlights that inter-multiplet $J$-mixing mainly acts to renormalize the effective Stevens coefficients and
the resulting low-energy spectrum, while leaving the overall level hierarchy intact.

Overall, we note that point-charge CEF calculations are approximate in nature (e.g., neglecting covalency and
orbital-dependent shielding) and should be treated as providing a qualitative estimate of the single-ion
anisotropy and level hierarchy rather than a parameter-free quantitative description. Within this level of
approximation, the calculation confirms that the low-temperature physics of NdMgAl$_{11}$O$_{19}$ (< $10$~K) is governed by an
isolated Kramers ground-state doublet (effective two-level system), and the resulting anisotropic $g$ tensor is in
good agreement with the experimental $g$ values extracted from X-band EPR and with those determined in the isostructural analogue NdZnAl$_{11}$O$_{19}$~\cite{cao2025}, supporting the identification of a
dominant easy-axis response along the crystallographic $c$ axis. A definitive determination of the full CEF level
scheme and absolute parameters requires direct spectroscopic probes (e.g., inelastic neutron scattering), which is
therefore recommended for future work.

\section{Angle-Dependent EPR Spectroscopy}\label{app:epr}

\begin{figure}[t]
\centering
\includegraphics[width=0.480\textwidth]{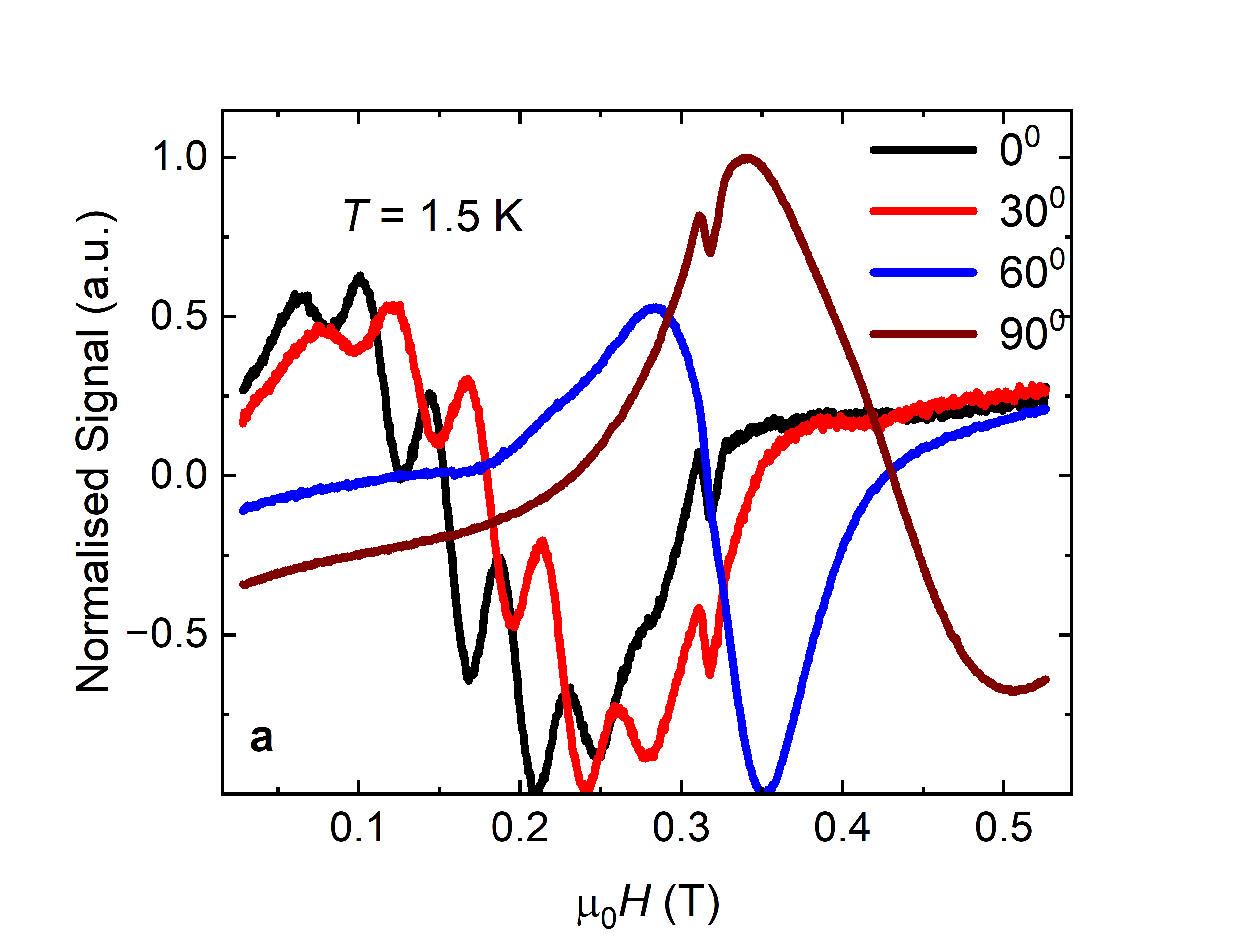}\hfill
\includegraphics[width=0.480\textwidth]{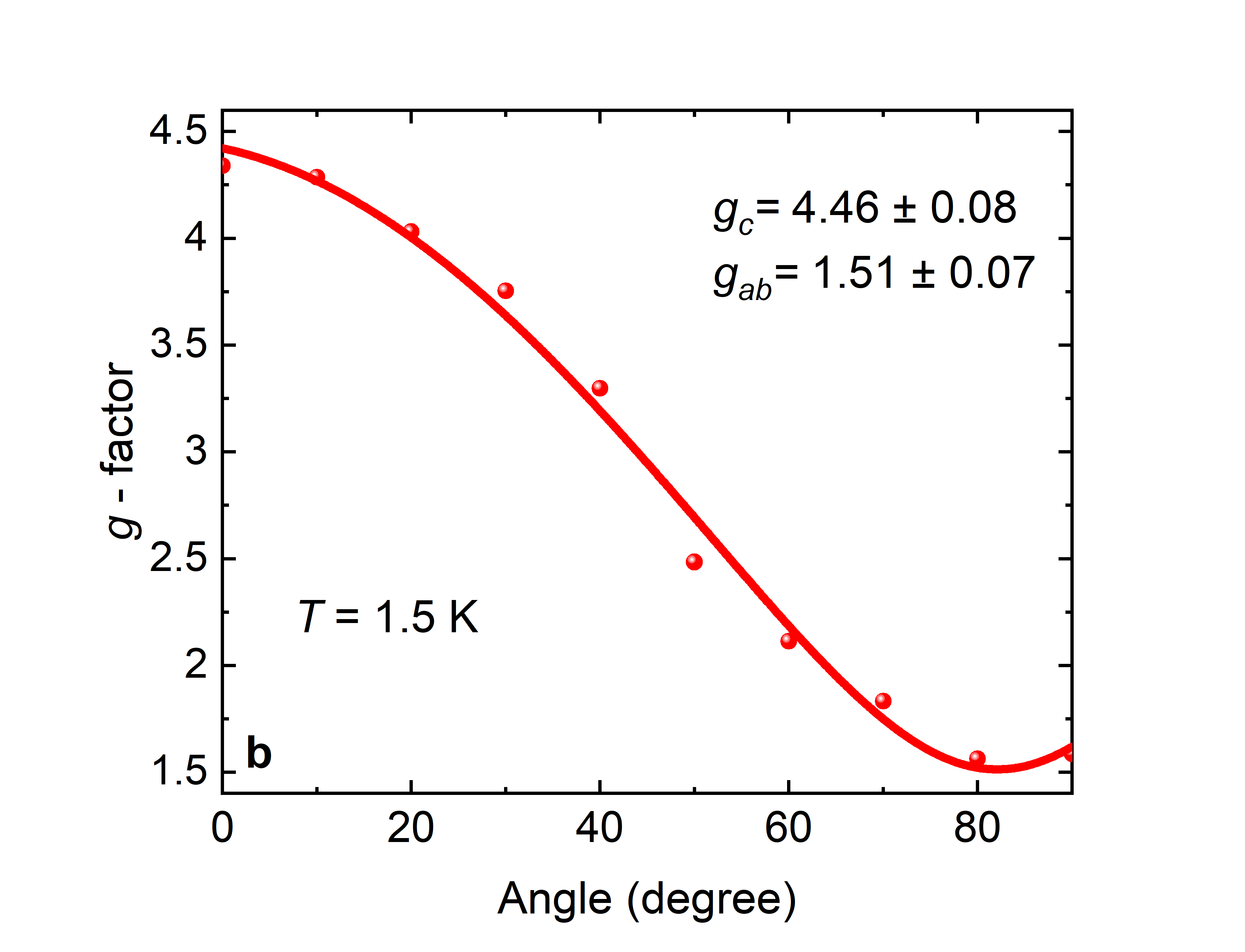}
\caption{
(a) Normalized X-band EPR spectra ($dI/dB$ vs.\ $B$, $\nu = 9.4$~GHz) of a NdMgAl$_{11}$O$_{19}$ 
single crystal at \SI{1.5}{\kelvin} for rotation angles $\theta = 0^\circ$, $30^\circ$, $60^\circ$, 
and $90^\circ$. Fine structure visible near $H \parallel c$ (\(\theta = 0^\circ\)) gradually fades 
upon rotation toward the hard plane ($H \parallel ab$).  
(b) Effective $g$-factor as a function of rotation angle $\theta$, extracted from the 
extrema–midpoint resonance field $B_{\mathrm{res}}$. 
The solid red curve represents the fit to the uniaxial angular-dependence model.
}

\label{figS2}
\end{figure}

Continuous-wave (CW) X-band electron paramagnetic resonance (EPR) measurements 
(\(\nu = 9.4~\text{GHz}\)) were performed on a single crystal of NdMgAl$_{11}$O$_{19}$ at 
\SI{1.5}{\kelvin}. The rotation angle \(\theta\) was defined such that \(\theta = 0^\circ\) corresponds 
to \(H \parallel c\) (easy axis), whereas \(\theta = 90^\circ\) corresponds to \(H \parallel ab\) 
(hard plane). Spectra were acquired in first-derivative mode using lock-in detection, yielding 
signals of the form \(dI/dB\), which were plotted as a function of magnetic field \(B\) 
(Fig.~\ref{figS2}(a)).

A linear background was subtracted using least-squares detrending. For visualization, the 
stacked spectra were interpolated onto a uniform magnetic-field grid and normalized to 
unit amplitude, while all quantitative analysis was performed on the raw
intensity data.

The resonance field \(B_{\mathrm{res}}\) was extracted using the standard extrema–midpoint method,
\begin{equation}
B_{\mathrm{res}} = \frac{B_{\max} + B_{\min}}{2},
\end{equation}
where \(B_{\max}\) and \(B_{\min}\) denote the field positions of the positive and negative 
derivative extrema, respectively~\citep{Weil2006,Bodziony2008JAC}. The corresponding peak-to-peak linewidth 
was defined as
\begin{equation}
\Delta B_{\mathrm{pp}} = B_{\min} - B_{\max}.
\end{equation}

The effective \(g\)-factor was obtained from the resonance condition
\begin{equation}
g = \frac{h\nu}{\mu_B B_{\mathrm{res}}},
\end{equation}
and used to construct the angular dependence \(g(\theta)\) shown in 
Fig.~\ref{figS1}(b).

The angular variation of $g(\theta)$ is consistent with strong uniaxial anisotropy for Nd$^{3+}$ in a trigonal crystal field. Symbols in the $g(\theta)$ plot denote experimental data points.  
At $\theta = 0^\circ$ ($\mathbf{H} \parallel c$), the spectrum exhibits a weak multi-hump fine structure that gradually disappears upon rotation toward the $ab$ plane. The fine structure is clearly visible near $0^\circ$, weakens at intermediate angles, and becomes unresolved above approximately $60^\circ$.

The angular dependence of the effective $g$-factor was fitted using the standard axial
anisotropy model (Fig.~\ref{figS2}(b))
\[
g(\theta) = \sqrt{ g_c^2 \cos^2(\theta - \theta_0) + g_{ab}^2 \sin^2(\theta - \theta_0) }.
\]
The fit yields $g_c = 4.46(8)$ and $g_{ab} = 1.51(7)$, revealing strong uniaxial 
anisotropy characteristic of Nd$^{3+}$ in trigonal coordination. The small offset angle 
$\theta_0 = -7.92(16)^\circ$ reflects a minor misalignment between the crystal’s 
$c$-axis and the rotation axis of the goniometer. The fit confirms that the angular variation of the 
resonance field is well described by an axial $g$-tensor, in excellent agreement with 
the point-charge CEF result ($g_{zz}=4.45$, $g_{xx}=g_{yy}=1.35$). A similar value has been reported for the Zn analogue NdZnAl$_{11}$O$_{19}$ ~\citep{cao2025}.

This fine structure along the $c$ axis originates primarily from spin–orbit-induced crystal-field mixing within the lowest Kramers doublet of Nd$^{3+}$, combined with unresolved hyperfine interactions ~\cite{aminov2013,Bodziony2008JAC}.  
The $4f^{3}$ ($J = 9/2$) ground multiplet splits into several Kramers doublets under the trigonal crystal field.  
A magnetic field applied along the symmetry axis admixes the two lowest doublets, producing nearby resonances with slightly different effective $g_c$ values.  
This mixing, together with hyperfine coupling, gives rise to the observed multi-hump structure at low angles and its progressive suppression upon tilting, when the quantization axis deviates from the crystal-field axis.  
Additional broadening may result from local strain or weak site inequivalence, although no fully resolved hyperfine multiplets are detected.

In summary, the EPR results reveal pronounced uniaxial anisotropy ($g_c \gg g_{ab}$) for Nd$^{3+}$ in trigonal coordination, with fine structure along the easy axis arising from spin–orbit-driven mixing in the Kramers manifold and enhanced by hyperfine coupling.  
The structure coalesces into a single line in the hard plane.  
\end{appendices}

\bibliographystyle{aipnum4-2}

\bibliography{main}

\end{document}